\documentclass[a4paper,11pt]{article}
\pdfoutput=1 

\usepackage{physics}
\usepackage{jheppub} 

\usepackage[T1]{fontenc} 
\usepackage[utf8x]{inputenc}
\usepackage{hyperref}
\usepackage[numbers]{natbib}
\usepackage{subcaption}
\usepackage{pgfplots}
\usepackage{comment}
\usetikzlibrary{angles,quotes}
\usepackage[normalem]{ulem}

\title{\boldmath Effects of background rotation and anisotropy in the holographic description of type-II superconductors

}

\author[a,b]{Jhony A. Herrera-Mendoza,}
\author[a]{Alfredo Herrera-Aguilar,}
\author[a,c]{Daniel F. Higuita-Borja,}
\author[d]{Julio A. Méndez-Zavaleta,}
\author[a]{Felipe Pérez-Rodríguez,}
\author[e]{Jia-Xin Yin}

\affiliation[a]{Instituto de F\'isica, Benem\'erita Universidad Aut\'onoma de Puebla, Edificio IF-1, Ciudad Universitaria, Puebla, Pue. 72570, México.}
\affiliation[b]{Escuela de F\'isica, Facultad de Ciencias, Universidad Nacional Aut\'onoma de Honduras,
Blvr. Suyapa, Tegucigalpa, Municipio del Distrito Central 11101, Honduras.} 
\affiliation[c]{Instituto de F\'isica, Universidad de Antioquia, Calle 70 No 52-21, Medell\'in, Colombia.}
\affiliation[d]{Facultad de Física, Universidad Veracruzana, Paseo No. 112, Desarrollo Habitacional Nuevo Xalapa, CP 91097, Xalapa-Enr\'{i}quez, México.}
\affiliation[e]{Department of Physics, Southern University of Science and Technology, Shenzhen, Guangdong 518055,
	China.}

\emailAdd{jhonyahm@gmail.com}
\emailAdd{aherrera@ifuap.buap.mx}
\emailAdd{dfhiguit@gmail.com}
\emailAdd{julmendez@uv.mx}
\emailAdd{fperez@ifuap.buap.mx}
\emailAdd{yinjx@sustech.edu.cn}

\abstract{The present work concerns the detailed construction of a holographic model for a type-II s-wave superconductor defined on a 5-dimensional anisotropic rotating black hole. We examine the role of rotation and anisotropy on the properties of the superconductor model focusing on the condensate and the AC conductivity, for which we obtain closed formulas, using both analytical and numerical methods. The results reveal that the rotation is responsible for the appearance of a peak and for introducing an exponentially vanishing behavior in the high-frequency limit of the real component of the AC conductivity. 
Such a behavior aligns with that observed in high-temperature superconductor models and experiments, where the peak and vanishing behavior result from quasiparticle damping, suggesting a relation between the {\it rotation of a black hole} and {\it quasiparticle damping effects} due to impurities or defects in a superconducting material. This relation supplements the holographic dictionary of the gravity/Condensed Matter Theory correspondence. 
In addition, we provide a detailed construction of the vortex lattice presented in  \cite{Herrera-Mendoza:2022whz} and study its behavior as a function of an external uniform magnetic field. Once again, it is shown that the vortex lattice can be continuously deformed along with a change in the vortex population by virtue of the magnetic field, providing a promising avenue for holographically modeling the vortex lattice deformations observed in experimental studies with superconducting materials. As a concrete example, we describe both the vortex lattice deformation and the increment of the vortex population under the action of an external magnetic field in the LiFeAs type-II superconductor. These effects supplement those previously found for the FeSe type-II superconductor studied in \cite{Herrera-Mendoza:2022whz}.}

\begin{document} 
\maketitle
\flushbottom
\section{Introduction}\label{sec:Intro}

Over the past years, the gauge/gravity duality has emerged as a powerful tool for investigating various phenomena within diverse physical contexts. Initially introduced as the AdS/CFT correspondence \cite{Maldacena:1997re}, this duality establishes a relation between a quantum field theory at strong coupling defined on a flat spacetime and a gravitational theory in a higher-dimensional spacetime. This approach has proven particularly useful for studying strongly coupled quantum systems, such as those found in condensed matter physics and nuclear physics, where traditional analytical techniques fail to provide insight. In this regard, notable applications include the holographic description of high temperature superconductors \cite{Hartnoll:2008vx} and superfluids \cite{Herzog:2008he}. From the condensed matter theory picture, strong-coupling superconductivity is understood as high-temperature superconductivity induced by a very strong phonon-mediated pairing interaction \cite{Rainer_1995}. Within this framework, typical energies of superconducting electrons
are similar to energies of lattice excitations that mediate the attractive interaction. Thus, superconductivity is sensitive to the dynamical properties of these excitations, which departs from the laws of weakly-coupled or low-temperature superconductivity. These physical properties motivate the search for dual gravitational descriptions capable of capturing such nontrivial dynamics.

These seminal works laid the groundwork for subsequent developments within the frameworks of holographic superconductivity and superfluidity. Notably, holographic models of superconductors and superfluids can emerge from a variety of gravitational setups supported by diverse matter content. These include not only asymptotically AdS backgrounds but also Lifshitz geometries \cite{Taylor:2008tg,Ayon-Beato:2009rgu,Ayon-Beato:2010vyw,Ayon-Beato:2019kmz,Bravo-Gaete:2020ftn,Bravo-Gaete:2021kgt}, which have proven valuable for modeling field theory systems with quantum critical behavior governed by non-relativistic anisotropic scaling between space and time \cite{Kachru:2008yh}. 
More concretely, holographic models of superconductors and superfluids have been extensively developed in asymptotically AdS backgrounds \cite{Hartnoll:2008vx,Hartnoll:2008kx,Hartnoll:2009sz,Herzog:2008he,Gangopadhyay:2012am,Gangopadhyay:2012np}. These studies include the effects of uniform external magnetic fields, which led to the holographic modeling of Abrikosov vortex lattices \cite{Nakano:2008xc,Albash:2008eh,Albash:2009iq,Albash:2009ix,Maeda:2009vf,Montull:2009fe,Adams:2012pj,Xia:2019eje,Srivastav:2023qof}. On the other hand, non-relativistic holographic superconductors have been investigated in the context of asymptotically Lifshitz spacetimes \cite{Bu:2012zzb,Lu:2013tza,Natsuume:2018yrg,Zhao:2013pva}. Interestingly, and of particular relevance to the present work, rotational effects have also been studied in both AdS \cite{Sonner:2009fk,Lin:2014tza,Srivastav:2019ixc} and Lifshitz scenarios \cite{Herrera-Mendoza:2022whz}.

This work develops a comprehensive holographic model of a type-II superconductor defined on a rotating, anisotropic black hole background. By incorporating a rotating background, our primary goal is to investigate the impact of rotation on the properties of a holographic superconductor, with special attention to the condensate phenomena and AC conductivity for which closed expressions were obtained. The results indicate the following  relationship 
\begin{equation*}
\begin{array}{ccccc}
\text{
the role of rotation in} & \qquad & & \qquad &
\text{
the quasiparticle damping effect}\\
\text{ 
\, an anisotropic black hole } & \qquad &
\Longleftrightarrow & \qquad &
\text{
\, induced by impurities or defects }\\
\text{
gravitational configuration} & \qquad & & \qquad &
\text{
in a superconducting material}
\end{array}
\end{equation*}
\noindent Additionally, we explore the formation of the vortex lattice and its deformation under an external magnetic field, a common feature of type-II superconductors. It is observed how infinitesimal variations of an external magnetic field influence the dynamics of vortices, changing their population and causing continuous deformations in the vortex lattice of type-II holographic superconductors. Remarkably, both of these physical effects are experimentally observed in the LiFeAs type-II superconductor under the action of an external magnetic field. 

\emph{The manuscript is organized as follows}: in Sec. II we introduce the holographic setup working in the so-called probe limit. Sec. III is devoted to studying the condensation of the scalar field using both analytical and numerical methods, while Sec. IV presents an analytical treatment of the AC conductivity supplemented with numerical computations. In addition, in Sec. V we investigate the effect of a uniform external magnetic field on the vortex lattice, uncovering how the field influences the arrangement of vortices. We conclude in Sec. VI with a discussion of our relevant findings and their implications.
\section{The setup}
\label{sec:RotatingConfig}
According to the gauge/gravity correspondence, we start by defining the higher dimensional gravity theory of interest. For our purposes let us consider the following action
\begin{equation}\label{eq:EMD2}
	S=\int{d^{d+2}x\sqrt{-g}\Big(\mathcal{L}_{\text{bg}}-\dfrac{1}{q^2}\mathcal{L}_{\text{matter}}\Big)},
\end{equation}
with background ($\mathcal{L}_{\text{bg}}$) and matter ($\mathcal{L}_{\text{matter}}$) Lagrangians given by
\begin{subequations}\label{eq:L_setup}
	\begin{align}
		\mathcal{L}_{\text{bg}} &= \dfrac{R-2\lambda}{2\kappa}-\dfrac{1}{2}\nabla_\mu\varphi\nabla^\mu\varphi-\dfrac{1}{4}e^{-b\varphi}\mathcal{F}_{\mu\nu}\mathcal{F}^{\mu\nu},\\
		\mathcal{L}_{\text{matter}}&=\dfrac{1}{4}F_{\mu\nu}F^{\mu\nu}+\dfrac{1}{\ell^2}\left(|D_\mu\Psi|^2+m^2|\Psi|^2\right).
	\end{align}
\end{subequations}
The background Lagrangian $\mathcal{L}_{\text{bg}}$ is defined by the Einstein-Hilbert action with cosmological constant $\lambda$, supplemented by a scalar dilaton field $\varphi$ and a gauge field strength $\mathcal{F}_{\mu \nu}$. While the matter Lagrangian $\mathcal{L}_{\text{matter}}$ is defined by a gauge field strength $F_{\mu \nu}$ and a complex scalar field $\Psi$,
 charged by the covariant derivative $D_\mu\equiv\nabla_\mu-iA_\mu$. In addition, we take the limit $q\rightarrow\infty$ wherein the Lagrangian $\mathcal{L}_{\text{bg}}$ can be considered decoupled from $\mathcal{L}_{\text{matter}}$, such that we can neglect the backreaction of $\mathcal{L}_{\text{matter}}$ and work the equations of motion associated to $\mathcal{L}_{\text{bg}}$ in order to look for a concrete gravitational solution with a manifest Lifshitz symmetry. 

Recall the rotating, anisotropic black hole solution for the $\mathcal{L}_{\text{bg}}$ sector obtained in \cite{Herrera-Aguilar:2021top}
\begin{align}\label{eq:Cov_Ansatz}
	ds^2=-\left(\dfrac{r}{\ell}\right)^{2z}f(r)\left(\Xi dt-\sum_{i=1}^{n}a_id\phi_i\right)^2 & +\dfrac{r^2}{\ell^4}\sum_{i=1}^{n}\left(a_i dt-\Xi \ell^2 d\phi_i\right)^2+\dfrac{dr^2}{\left(\dfrac{r}{\ell}\right)^2f(r)}\nonumber\\
	& -\dfrac{r^2}{\ell^2}\sum_{i<j}^{n}\left(a_id\phi_j-a_jd\phi_i\right)^2+\left(\dfrac{r}{\ell}\right)^2d\vec{y}{}^2_{(d-n)}, 
\end{align}
with 
\begin{equation}
	\label{eq:TaylorSols}
	\begin{split}
		&f(r) =1-\qty(\dfrac{r_{h}}{r})^{z+d},\qquad \qquad \qquad \qquad \quad 
		e^{-b\varphi}  =\kappa\dfrac{Q^2\ell^2}{(z-1)(z+d)}\left(\dfrac{\ell}{r}\right)^{2d},\\
		&\mathcal{F}_{tr}  =\Xi\dfrac{(z-1)(z+d)}{\kappa Q\ell^2}\left(\dfrac{r}{\ell}\right)^{d+z-1},\qquad
		\mathcal{F}_{\phi_ir}  =-a_i\dfrac{(z-1)(z+d)}{\kappa Q\ell^2}\left(\dfrac{r}{\ell}\right)^{d+z-1},\\
		&\Lambda  =-\dfrac{(z+d)(z+d-1)}{2\ell^2},\qquad \qquad  \qquad 
		b  =2\sqrt{\dfrac{\kappa d}{z-1}}.
	\end{split}
\end{equation}
This gravitational configuration describes a rotating Lifshitz-like black hole in \(d+2\) dimensions, supported by a dilaton field, a Maxwell field with electric and magnetic components, and a negative cosmological constant. It includes up to $n = [(d+1)/2]$ independent rotation planes, each associated with a parameter $a_i$, leading to anisotropy and rotational effects encoded in the metric through $\Xi = \sqrt{1 + \sum_{i=1}^n a_i^2/\ell^2}$. The spacetime exhibits Lifshitz scaling with dynamical exponent \(z\), a black hole horizon at $r = r_h$, and flat transverse directions described by the Euclidean metric $d\vec{y}{}^2_{(d-n)} = dy_kdy^k$, making it a suitable background for holographic models of strongly correlated systems with anisotropic and rotating features. The temperature $T$ and angular momentum $\mathcal{J}_i$ of this configuration are respectively given by
\begin{equation}\label{eq:BBTemperature}
\quad T=\dfrac{1}{4\pi}\dfrac{(z+d-2)r_h^z}{\ell^{z+1} \Xi}, \quad \quad \mathcal{J}_i=\dfrac{a_i \Xi(d-z) V_{(d-2)} r_h^{z+d-2}}{2\kappa \ell^{z+[d/2]}}.
\end{equation}
See \cite{Herrera-Aguilar:2021top} for a deeper analysis concerning the physical properties of this background.

For the purposes of this work, we consider a particular case of spacetime \eqref{eq:Cov_Ansatz} in five dimensions, assuming only one non-trivial rotation parameter (\(a_1 = a\), \(a_2 = 0\)). In this case, the metric reduces to
\begin{equation}\label{eq:Cov_Ansatz5D}
	ds^2=-\left(\dfrac{r}{\ell}\right)^{2z}f(r)\left(\Xi dt-ad\phi\right)^2+\dfrac{r^2}{\ell^4}\left(a dt-\Xi \ell^2 d\phi\right)^2+\dfrac{dr^2}{\left(\dfrac{r}{\ell}\right)^2f(r)}+\left(\dfrac{r}{\ell}\right)^2d\vec{y}{}^2, 
\end{equation}
with $d\vec{y}{}^2=dx^2+dy^2$ and $f(r)=1-\left(\dfrac{r_h}{r}\right)^{z+3}$. 

\subsection{The decoupling limit}
As we discussed before, in the limit $q\rightarrow\infty$, the feedback or backreaction of the matter fields in $\mathcal{L}_{\text{matter}}$ is suppressed on the bulk fields of $\mathcal{L}_{\text{bg}}$. In this manner, we obtain the following set of field equations on the now fixed background \eqref{eq:Cov_Ansatz5D}, i.e.
\begin{subequations}\label{eq:GenFieldEqns_rot}
	\begin{align}
		\nabla_\mu F^{\mu\nu}-\dfrac{1}{\ell^2}\left(2A^\nu|\Psi|^2-i\Psi\nabla^\nu\overline{\Psi}+i\overline{\Psi}\nabla^\nu\Psi\right) & =0,\\
		\nabla^{\mu}\nabla_{\mu}\Psi-2iA^{\mu}\nabla_{\mu}\Psi-i\Psi\nabla_{\mu}A^{\mu}-\left(m^2+A_{\mu}A^{\mu}\right)\Psi & =0.
	\end{align}
\end{subequations}
Next, we take the most general stationary and axisymmetric ans\"{a}tze for the matter fields compatible with the metric structure \eqref{eq:Cov_Ansatz5D},
\begin{equation}
	\Psi=\Psi(r,\vec{y}), \quad A_t=A_t(r,\vec{y}), \quad A_\phi=A_\phi(r,\vec{y}), \quad A_x=A_x(r,\vec{y}), \quad A_y=A_y(r,\vec{y}).
\end{equation}
For simplicity, we redefine the holographic coordinate as $u=r_h/r$, in which the boundary and horizon are located at $u=0$ and $u=1$, respectively.
By taking into account the previous information, the field equations \eqref{eq:GenFieldEqns_rot} reduce to the following coupled nonlinear system of partial differential equations.\\
The scalar equation
\begin{align}\label{eq:gen_scalar_eq}
	\partial^2_{uu}\Psi&+\left[\dfrac{\partial_uf}{f}-\dfrac{(z+2)}{u}\right]\partial_u\Psi+\dfrac{\ell^4}{r_h^2f}\left(\partial^2_{xx}+\partial^2_{yy}\right)\Psi -\dfrac{i\ell^4}{r_h^2f}[(\partial_xA_x+\partial_yA_y)\Psi+2A_x\partial_x\Psi\nonumber\\ &+2A_y\partial_y\Psi]
	-\left\{\dfrac{l^2}{u^2f}\left[m^2+\left(\ell^2(A_x^2+A_y^2)+(aA_t+\Xi A_\phi)^2\right)\qty(\dfrac{u}{r_h})^2\right]\right.\nonumber \\ & \left.-\qty(\dfrac{\ell u}{r_h})^{2(z-1)}\dfrac{(\ell^2\Xi A_t+aA_\phi)^2}{r_h^2f^2}\right\}\Psi=0,
\end{align}
and the Maxwell equations
\begin{subequations}\label{eq:FieldEqns}
	\begin{align}
		\partial^2_{uu}A_t&+\left[\dfrac{2a^2(z-1)+\ell^2(z-2)}{u}-a^2\dfrac{\partial_uf}{f}\right]\dfrac{\partial_uA_t}{\ell^2}+\left[\dfrac{2a\Xi(z-1)}{u}-a\Xi\dfrac{\partial_uf}{f}\right]\dfrac{\partial_uA_\phi}{\ell^2} \nonumber\\
		&+\dfrac{\ell^4}{r_h^2f}\left(\partial^2_{xx}+\partial^2_{yy}\right)A_t-\dfrac{2}{u^2f}|\Psi|^2A_t =0,\label{eq:gen_At_eq}\\
		\partial^2_{uu}A_\phi &-\left[\dfrac{2a^2(z-1)+\ell^2z}{u}-\ell^2\Xi^2\dfrac{\partial_uf}{f}\right]\dfrac{\partial_uA_\phi}{\ell^2}-\left[\dfrac{2a\Xi(z-1)}{u}-a\Xi\dfrac{\partial_uf}{f}\right]\partial_uA_t  \nonumber\\
		&+\dfrac{\ell^4}{r_h^2f}\left(\partial^2_{xx}+\partial^2_{yy}\right)A_\phi-\dfrac{2}{u^2f}|\Psi|^2A_\phi  =0,\label{eq:gen_Aphi_eq}\\
		\partial^2_{uu}A_y&+\left(\dfrac{\partial_uf}{f}-\dfrac{z}{u}\right)\partial_uA_y+\dfrac{\ell^4}{r_h^2f}\left(\partial^2_{xx}A_y-\partial^2_{xy}A_x\right)+\dfrac{2}{u^2f}\mathfrak{Im}(\bar{\Psi}\partial_y\Psi)\nonumber \\&-\dfrac{2}{u^2f}|\Psi|^2A_y  =0,\label{eq:gen_Ay_eq}\\
		\partial^2_{uu}A_x &+\left(\dfrac{\partial_uf}{f}-\dfrac{z}{u}\right)\partial_uA_x+\dfrac{\ell^4}{r_h^2f}\left(\partial^2_{yy}A_x-\partial^2_{xy}A_y\right)+\dfrac{2}{u^2f}\mathfrak{Im}(\bar{\Psi}\partial_x\Psi)\nonumber \\&-\dfrac{2}{u^2f}|\Psi|^2A_x  =0,\label{eq:gen_Ax_eq}\\
		\partial^2_{ux}A_x&+\partial^2_{uy}A_y-\dfrac{r_h^2}{\ell^4u^2}2\mathfrak{Im}(\bar{\Psi}\partial_u\Psi)  =0\label{eq:gen_Ax_Ay_eq},
	\end{align}
\end{subequations}
We notice a strong coupling between the vector and scalar sectors, even when disregarding the background interaction. While finding an exact solution remains unattainable, the subsequent section delves into an approach that facilitates the characterization of the physics associated with these fields.  
\section{Condensation of the scalar field}\label{sec:Condensate}
Holographic superconductivity is characterized by the presence of a scalar field acting as the order parameter that condensates below some critical temperature. Here, we want to explore the condensation of the scalar field $\Psi$ in our field configuration. To achieve this, we consider real scalar and gauge fields with the simplest ans\"{a}tze consistent with the equations \eqref{eq:gen_scalar_eq} and \eqref{eq:FieldEqns}, i.e.
\begin{equation}
	\Psi=\psi(u), \qquad A_t=A_t(u), \qquad A_\phi=A_\phi(u), \qquad A_x=0, \qquad A_y=0.
\end{equation}
In such a scenario the system of equations \eqref{eq:gen_scalar_eq}-\eqref{eq:FieldEqns} reduces to
\begin{subequations}\label{eq:CondeFieldEqns}
	\begin{align}
		&A_t''+\left[\dfrac{2a^2(z-1)+\ell^2(z-2)}{u}-a^2\dfrac{f'}{f}\right]\dfrac{A_t'}{\ell^2}+\left[\dfrac{2a\Xi(z-1)}{u}-a\Xi\dfrac{f'}{f}\right]\dfrac{A_\phi'}{\ell^2}\nonumber\\&\quad \,-\dfrac{2}{u^2f}\psi^2A_t  =0,\\
		&A_\phi''-\left[\dfrac{2a^2(z-1)+\ell^2z}{u}-\ell^2\Xi^2\dfrac{f'}{f}\right]\dfrac{A_\phi'}{\ell^2}-\left[\dfrac{2a\Xi(z-1)}{u}-a\Xi\dfrac{f'}{f}\right]A_t'-\dfrac{2}{u^2f}\psi^2A_\phi  =0,\\
		&\psi''+\left[\dfrac{f'}{f}-\dfrac{(z+2)}{u}\right]\psi'-\Biggl\{\dfrac{\ell^2\left(aA_t+\Xi A_\phi\right)^2}{r_h^2f}-\qty(\dfrac{\ell u}{r_h})^{2(z-1)}\dfrac{(\ell^2\Xi A_t+aA_\phi)^2}{r_h^2f^2}\nonumber\\  &\quad \, +\dfrac{\ell^2m^2}{u^2f}\Biggr\}\psi  =0.
	\end{align}
\end{subequations}
In order to explore the condensation phenomenon, we need to solve this system at the asymptotic boundary ($u=0$). In this regard, consistent with the holographic prescription, it is simple to show that the system adopts the following boundary solutions
\begin{equation}\label{eq.BoundarySol}
	\psi=J_{-}u^{\Delta_{-}}+J_{+}u^{\Delta_{+}}, \quad A_t=\mu-\rho\qty(\dfrac{u}{r_h})^{3-z}, \quad A_\phi=\nu-\zeta\qty(\dfrac{u}{r_h})^{3-z}, 
\end{equation}
provided $z\leq3$, where $J_{-}$ ($J_{+}$) can be identified as the source (the expectation value) of the scalar operator $\mathcal{O}$, with $\Delta_\pm=[(z+3)\pm\sqrt{(z+3)^2+4\ell^2m^2}]/2$ the corresponding scaling dimensions. To ensure the existence of normalizable modes near the asymptotic boundary, the mass of the scalar field must satisfy the Breitenlohner-Freedman (BF) bound for massive scalars in a Lifshitz background, given by $m^2\ell^2 \ge -\frac{(z+3)^2}{4}$. Additionally, $\mu$ and $\nu$ are interpreted as potentials in the dual field theory, while $\rho$ and $\zeta$ represent the charge density and the current density, respectively. In this specific field configuration, these densities are related by $\zeta=-\dfrac{a}{\Xi}\rho$.

The holographic dictionary facilitates the relation $\expval{\mathcal{O}}=\sqrt{2}r_h^\Delta J$, in which $J$ is a function of the background parameters.
\begin{figure}[tbp]
	\centering
\includegraphics[width=.495\textwidth]{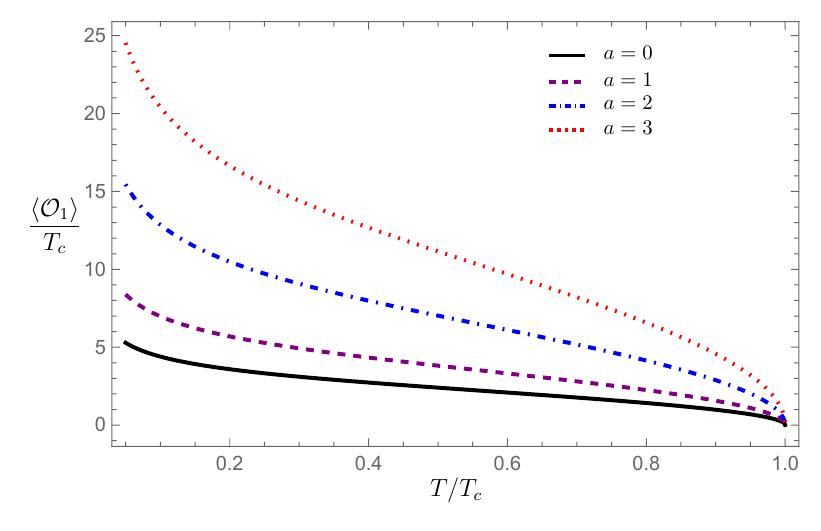}
\includegraphics[width=.495\textwidth]{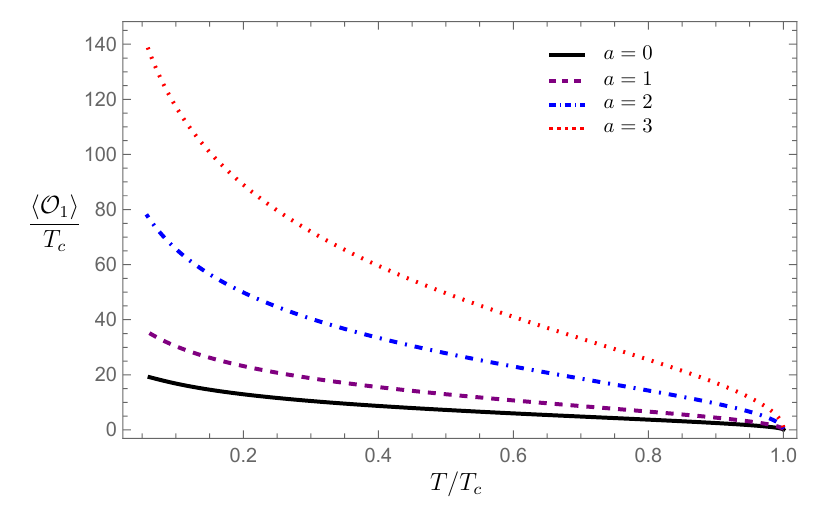} 
\caption{Profiles of the condensation operator taking different values of the rotation parameter and considering the boundary condition $J_{+} = 0$, $J_{-} \sim \expval{\mathcal{O}_1}$, and fixed parameters $\ell = 1$, $z=2$ and $\ell^2 m^2=-4$. The left panel shows the condensation of the dual operator derived via the matching method (equation \eqref{eq:ope_mm}), taking $u_m=0.8$. The right panel displays numerical results from solving \eqref{eq:CondeFieldEqns} using the shooting method.}
\label{fig:Condensate}
\end{figure}
In Appendix \ref{ape:A}, we present a detailed analytic derivation for $J$ (and $\expval{\mathcal{O}}$) in terms of the black hole parameters, using the matching method. This leads to the following expression for the expectation value of the order parameter:
\begin{align}\label{eq:ope_mm}
	\expval{\mathcal{O}}=\qty(\dfrac{4\pi\ell^{z+1}\Xi T_c}{z+3})^{\Delta/z}\qty(\dfrac{T}{T_c})^{(2\Delta-3)/2z} & \dfrac{u_m^{1-\Delta}[\ell^2m^2(1-u_m)+2(z+3)]}{\Delta(1-u_m)+2u_m}\times\nonumber\\
	& \sqrt{\dfrac{1+(z-2)(1-u_m)}{(z+3)(1-u_m)}\qty[1-\qty(\dfrac{T}{T_c})^{3/z}]}.
\end{align}
Here, $u_m$ is an intermediate bulk point introduced by the matching method, where the near-horizon and boundary expansions are smoothly joined. Moreover, the critical temperature $T_{c}$ is determined by imposing the condition $\expval{\mathcal{O}}=0$, yielding
\begin{equation}
	T_c=\dfrac{\ell^{z(z-3)/3-1}(z+3)}{4\pi\Xi^{1+z/3}}\qty{\dfrac{(3-z)\rho u_m^{2-z}}{\gamma[(z-2)(1-u_m)+1]}}^{z/3}.
\end{equation}
The expression~\eqref{eq:ope_mm} can be viewed as a holographic generalization of the Ginzburg–Landau order parameter when $ z \to 3 $ and $ \Delta \to 3/2 $, and it reproduces the characteristic scaling of the BCS gap near $ T_c$ for $ z \to 1 $ with the same scaling dimension. For general values of the dynamical exponent $ z $ and the scaling dimension $ \Delta $, it describes novel critical regimes that go beyond conventional superconducting models.

Figure \ref{fig:Condensate} depicts the condensation of the dual operator $\expval{\mathcal{O}}$ from two different perspectives: analytical (matching method) and numerical (shooting method). While both methods measure the same quantity, there are significant variations between them, as observed in the plots. However, it is not surprising that they do not yield identical results. Recall that the matching method involves an additional parameter, $u_m$, whose selection strongly influences the results for $\expval{\mathcal{O}}$. Hence, it only offers a qualitative explanation of the phenomenon. In contrast, the shooting method provides a more accurate depiction as it does not introduce any additional parameters and aligns with the anticipated outcome of the matching method. In other words, both figures exhibit the impact of varying the rotation parameter, which only scales the amplitude of the expectation value for the condensation operator as the rotation increases. 

\section{The AC conductivity}
\label{sec:Conductivities}
The AC conductivity is a fundamental quantity used to characterize a superconductor, as it captures the response of the system to external electric perturbations. In our holographic setup, this can be analyzed by introducing a gauge field perturbation within the bulk dynamics and solving the resulting linearized Maxwell equations. Specifically, we consider a perturbation in the $x$-component of the gauge field, $\delta A_x$, with a harmonic time dependence: $\delta A_{x} = A_{x}(u) e^{-i \omega t}$. Thus, after substituting this ansatz into \eqref{eq:GenFieldEqns_rot}, we obtain
\begin{equation}\label{eq:maxeq_Ax}
	A_x''+\qty(\frac{f'}{f}-\frac{z}{u})A_{x}'+\Biggl[\qty(\frac{\ell u}{r_{h}})^{2z}\frac{\omega^2\ell^2 \Xi^2}{f}-\Biggl(\frac{\ell^2 a^2 \omega^2 u^2}{r_{h}^2}+2 \psi^2 \Biggr)\Biggr]\frac{A_{x}}{u^2 f} =0.
\end{equation}
To solve this equation, we impose ingoing wave boundary conditions at the horizon
\begin{equation}\label{eq:AxH}
	A_{x} = (u-1)^{-i\frac{\ell^{z+1}}{(z+3)r_h^z}\omega}\Bigl[1+A_{x1}(u-1)+A_{x2}(u-1)^2+\cdots\Bigr],
\end{equation}
while the behavior at the asymptotic boundary is given by
\begin{equation}\label{eq:AxB}
	A_{x} = A_{x}^{(0)} + A_{x}^{(z+1)}\qty(\frac{u}{r_{h}})^{z+1}+\cdots,
\end{equation}
where, from the standard holographic dictionary, we interpret $A_{x}^{(0)}$ as the source of a conjugated current $J_{x}$, while $A_{x}^{(z+1)}$ is related to its expectation value, $\expval{J_{x}} \sim A_{x}^{(z+1)}$. Applying the Ohm's law, we thus obtain an expression for the AC conductivity 
\begin{equation}\label{eq:Ohms_law}
	\sigma(\omega) = \frac{\expval{J_{x}}}{E_x} = -\dfrac{(z+1)i}{\omega}\dfrac{ A_{x}^{(z+1)}}{ A_{x}^{(0)}}.
\end{equation}
In the remainder of this section, we focus in analyzing the frequency and angular momentum dependence of the conductivity. For this purpose, we conduct an analytical study of the Maxwell perturbation \eqref{eq:maxeq_Ax} considering the near-boundary limit of the background fields. We carry out this analysis by approximating $f \approx 1$ and $\psi \approx \expval{\mathcal{O}} u^{\Delta}/\sqrt{2}$, yielding
\begin{equation}\label{eq:maxeq_Ax_2}
	A_x''-\frac{z}{u}A_{x}'+\Biggl[\left(\qty(\frac{\ell u}{r_{h}})^{2z}\dfrac{\ell^2 \Xi^2}{u^2}-\frac{\ell^2 a^2}{r_{h}^2}\right)\omega^2- \dfrac{2\expval{\mathcal{O}}^2}{r_h^2} \left(\frac{u}{r_h}\right)^{2(\Delta-1)}  \Biggr]A_{x} =0.
\end{equation}
Analytical solutions to this equation are only obtainable for specific values of the parameters $\Delta$ and $z$.
In what follows, we consider the case $\Delta = 1$, which corresponds to a scalar mass squared of $m^2 = -(z+2)/\ell^2$. This value is consistent with the BF bound of Lifshitz spacetimes, ensuring the stability of the scalar sector. We present two specific examples where equation \eqref{eq:maxeq_Ax_2} admits analytic solutions. These results are further supported by numerical computations, in which we solve the full equation \eqref{eq:maxeq_Ax} subject to the boundary conditions \eqref{eq:AxH} and \eqref{eq:AxB}.

\subsection{Isotropic case: \texorpdfstring{$\Delta =1$}{text} and \texorpdfstring{$z=1$}{text}}
In this case, the equation \eqref{eq:maxeq_Ax_2} admits two independent solutions, which can be expressed in terms of Bessel functions of the first, $J_{1}$, and second kind, $Y_{1}$. Namely,
\begin{equation}\label{eq:SolAx1}
A_{x}(u)=c_0\, u\, J_1\left(i\, h(\omega)\, u\right)+  c_1\, u\, Y_1\left(-i\, h(\omega)\, u\right),
\end{equation}
where we define the auxiliary function
\begin{equation}\label{eq:def_h}
h(\omega)\equiv \dfrac{\sqrt{\expval{\mathcal{O}}^2-\ell^4 \omega ^2}}{r_{h}}.    
\end{equation}
Before proceeding, it is convenient to impose the following relation between the integration constants
\begin{equation}\label{eq:c0c1_1}
\dfrac{c_{0}}{c_{1}}=\dfrac{i}{\pi}\left[\pi H(\omega-\omega_g)+(1-2 \gamma_{E})i-\dfrac{180}{h^2(\omega)}+\frac{i}{2}\, \ln \left(\dfrac{h^2(\omega)}{4} \right)-\pi\right],
\end{equation}
in which $H(\omega-\omega_g)$ denotes the Heaviside step function, and $\gamma_{E}$ is the Euler-Mascheroni constant. The asymptotic analysis of the solution \eqref{eq:SolAx1}, combined with the relation \eqref{eq:Ohms_law}, yields the following expression for the conductivity 
\begin{equation}\label{eq:sigma1}
 \sigma(\omega) =  -\frac{{h^2(\omega)}}{\omega}\left(\pi H(\omega-\omega_g) +\dfrac{i}{2} \ln \left(\dfrac{h^2(\omega)}{4} \right)-\, \dfrac{180\, i}{h^2(\omega)} \right).
\end{equation}
A closer inspection of this relation uncovers the structure of real and imaginary parts of the conductivity, leading to
\begin{equation}\label{eq:SigmaRe_Analyticalc1} 
\text{Re}[\sigma] = \begin{cases} 
     \quad  \infty \qquad & \quad \omega = 0, \\
      \quad 0\qquad & 0<\omega \leq \omega_{g}, \\
      -\dfrac{3\pi}{4}\dfrac{h^2(\omega) }{\omega} &\quad \omega> \omega_{g}, \\
   \end{cases}
\end{equation}
\begin{equation}\label{eq:SigmaIm_Analyticalc1} 
\text{Im}[\sigma] = \begin{cases} 
      -\dfrac{{h^2(\omega)}}{2\omega}\ln \left(\dfrac{h^2(\omega)}{4}\right)+\dfrac{180}{\omega} \qquad & 0<\omega < \omega_{g}, \\
       -\dfrac{ {h^2(\omega)}}{2\omega} \ln\left(\dfrac{\abs{h^2(\omega)}}{4}\right)+\dfrac{180}{\omega} \qquad & \quad \omega> \omega_{g}. \\
   \end{cases}
\end{equation}
Here, $\omega_{g} = \expval{\mathcal{O}}/\ell^2$ denotes a gap frequency determined by the function $h(\omega)$ below which the real part of the conductivity vanishes---except at $\omega=0$, where a delta function appears.

\begin{figure}[tbp]
	\centering
\includegraphics[width=0.49\textwidth]{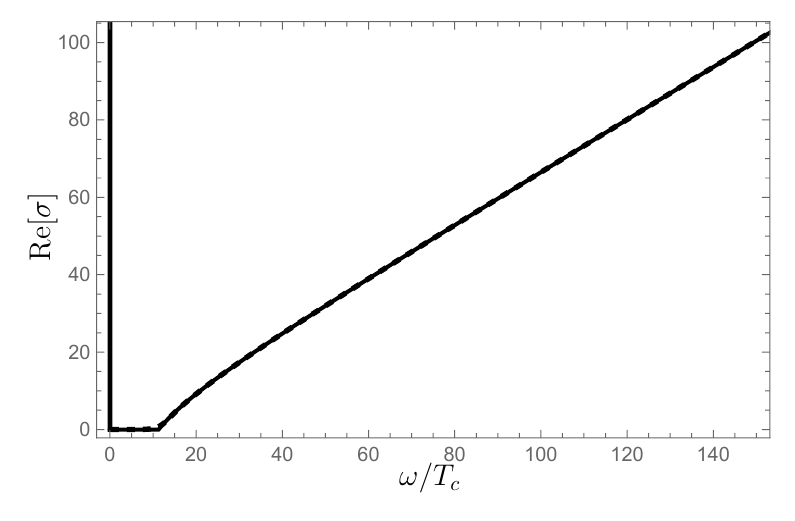}
\includegraphics[width=0.495\textwidth]{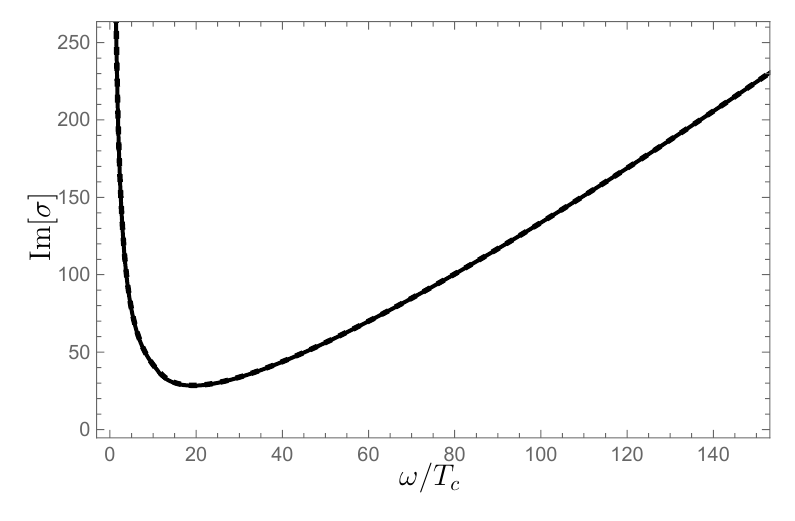}   
	\caption{Real (left) and imaginary (right) parts of the conductivity as functions of the frequency. The solid curves represent the analytical results, while the dashed ones correspond to numerical computations. These plots were generated at $T\approx 0.2\,T_c$, with parameters $\Delta =1$, $z=1$, $\ell =1$ and $r_h=1$.}
	\label{fig:case1}
\end{figure}

In Figure \ref{fig:case1}, we show the behavior of the real and imaginary parts of the conductivity defined in the previous relations \eqref{eq:SigmaRe_Analyticalc1}-\eqref{eq:SigmaIm_Analyticalc1}. The observed features exhibits recurring characteristics of those reported in previous holographic superconductor models. Regarding the real part of the conductivity, the presence of a delta function at $\omega =0$, indicates the emergence of a DC current and signals the onset of the superconducting state. Moreover, the existence of a gap frequency $\omega_{g}$, proportional to the order parameter $\expval{\mathcal{O}}$, is also a common property in holographic superconductors. However, it has been argued that holographic superconductors are not truly gapped; instead, the apparent gap arises from decoupling the matter sector from gravity, a consequence of employing the probe limit approximation. For an in-depth discussion, we refer the reader to \cite{Hartnoll:2009sz, Horowitz:2009ij, Horowitz:2010gk}.
For frequencies above the gap frequency $\omega > \omega_{g}$, the real part of the conductivity follows a monotonic increase approximately given by $\text{Re}[\sigma] \sim -h^2(\omega)/\omega$, following a linear tendency $\text{Re}[\sigma] \sim \omega$ at sufficiently higher frequencies. 

Turning our attention to the imaginary component \eqref{eq:SigmaIm_Analyticalc1}, we observe a minimum occurring very close to $\omega_{g}$. Below this minimum, the imaginary part exhibits a sharp increase characterized by $\text{Im}[\sigma] \sim 1/\omega$. Conversely, at sufficiently higher frequencies $\omega \gg \omega_{g}$, the behavior is given by $\text{Im}[\sigma] \sim \omega \ln \left(\ell^2 \omega/2r_h \right)$. These features are illustrated in Figure \ref{fig:case1}, where we highlight the excellent agreement between our analytical predictions (solid lines) and numerical results (dashed lines).

Lastly, it is important to note that we have not addressed the rotation of the background in the previous scenario. The reason for this is that neither of the expressions, \eqref{eq:SigmaRe_Analyticalc1} or \eqref{eq:SigmaIm_Analyticalc1}, possess information on the rotation parameter $a$. This observation results evident from the effective equation \eqref{eq:maxeq_Ax_2}, where the absence of anisotropy ($z=1$) leads to the cancellation of terms associated with rotation, rendering no residual information on this parameter. Consequently, we infer that anisotropy ($z\neq 1$) is an essential ingredient to capture the impact of rotation on the AC conductivity, at least within the scope of our holographic model.
\subsection{Anisotropic case: \texorpdfstring{$\Delta=1$}{text} and \texorpdfstring{$z=2$}{text}}
\begin{figure}[tbp]
	\centering
\includegraphics[width=0.495\textwidth]{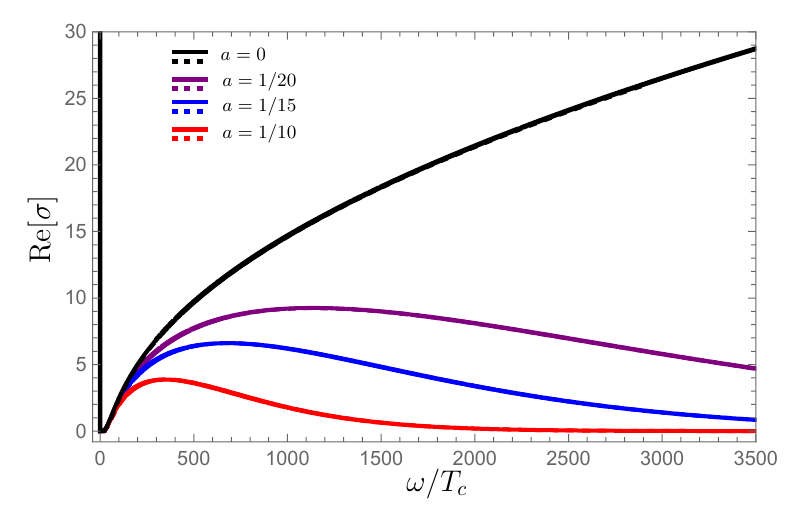}  
\includegraphics[width=0.495\textwidth]{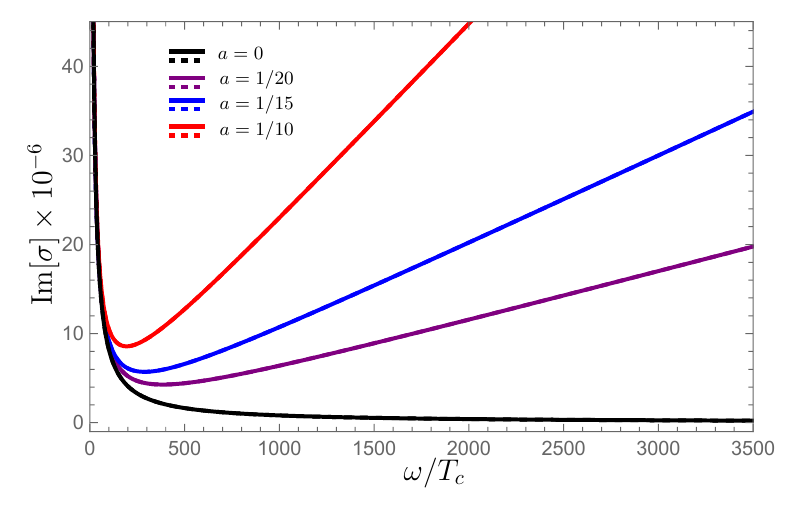}    
\caption{Portraits of the real (left) and imaginary (right) components of the conductivity as a function of the frequency. Note that the analytic curves (solid lines) show an excellent agreement with numerical ones (dashed lines). The plots were generated at $T\approx 0.061\,T_c$, considering $\Delta=2$, $z=2$, $\ell =1$ and $r_h=1$.}
\label{fig:case2}
\end{figure}

In this scenario, the solution can be written in terms of the confluent hypergeometric functions of the first kind, $M$, and second kind, $U$,
\begin{align}\label{eq:SolAx2}
A_{x}(u)=&e^{-i\frac{ \ell^3 \Xi \omega }{2 r_h^2}\, u^2} u^3 \Biggl[c_0\, M\left(\frac{5}{4}-i g(\omega,a),\, \dfrac{5}{2},\, i \dfrac{\Xi \ell^3 \omega }{r_h^2}\, u^2\right)\nonumber\\ &+ c_1\,U\left(\frac{5}{4}-i g(\omega,a),\, \dfrac{5}{2},\, i \dfrac{\Xi \ell^3 \omega }{r_h^2}\, u^2\right)\Biggr],
\end{align}
where we define
\begin{equation}\label{eq:def_g}
g(\omega,a) \equiv \dfrac{1}{4\ell \Xi }\left(a^2 \omega+\dfrac{\expval{\mathcal{O}}^2}{\ell^2\omega} \right).
\end{equation}
 As in the preceding case, we can obtain an analytic expression for the conductivity by employing the relation \eqref{eq:Ohms_law}. To accomplish this, we find it convenient to introduce the following identification 
\begin{align}\label{eq:c0c1_2}
\dfrac{c_{0}}{c_{1}}=&-\dfrac{\sqrt{\pi} e^{\pi g(\omega,a)}\left(1-i\,e^{2\pi g(\omega,a)} \right)}{\left(e^{4 \pi  g(\omega,a)}+1\right)\Gamma\left(-\frac{1}{4}-i\, g(\omega,a) \right)}\Biggl( \mathcal{F}(a)\,e^{-\pi g(\omega,a)}+\dfrac{4}{3} i\, e^{\pi g(\omega,a)}\nonumber\\& +i\,\mathcal{G}(a,\omega) \left(\dfrac{r_{h}^2}{\ell^3 \Xi}\right)^{3/2}\dfrac{\abs{\Gamma\left(-\frac{1}{4} +i\, g(\omega,a)\right)}^2}{2\pi(z+1)\omega^{1/2}\sech(2\pi g(\omega,a))}\Biggr),
\end{align}
in which the functions $\mathcal{F}(a)$ and $\mathcal{G}(a,\omega)$ adopt respectively the form
\begin{subequations}
\begin{align}
\mathcal{F}(a)&=\frac{711}{400} + \qty(\dfrac{a}{\ell})\exp[-\qty(\dfrac{a}{\ell})^2] \Biggl( \frac{3}{32} - \frac{464}{249} \qty(\dfrac{a}{\ell}) + \frac{829}{100} \qty(\dfrac{a}{\ell})^2 - \frac{22721}{1000} \qty(\dfrac{a}{\ell})^3 \nonumber 
\\ 
&\quad+ \frac{6497}{250} \qty(\dfrac{a}{\ell})^4 \Biggr), \label{eq:func1} \\ 
\mathcal{G}(a,\omega)&=  3  \ell \Biggl[ 6.61\times 10^{4} \left(\dfrac{a}{\ell}\right)^4+ 1.17\times 10^{4} \left(\dfrac{a}{\ell}\right)^2-763\left(\dfrac{a}{\ell}\right) + \dfrac{3}{2} +\dfrac{6.61\times10^{6}}{\ell \omega} \left(\dfrac{\expval{\mathcal{O}}}{\ell}\right)^2  \nonumber\\
&\quad+ 3.33 \times 10^6 \left(\dfrac{a}{\ell}\right)^2 \ell  \omega \Biggr]. \label{eq:func2}
\end{align}
\end{subequations}
Altogether, incorporating the definitions above, we arrive at the following analytic expression for the conductivity
\begin{equation}\label{eq:sigma2}
\sigma(\omega) =18\pi \left(\dfrac{\ell^3 \Xi}{r_{h}^2}\right)^{3/2}\left(\mathcal{F}(a)-\dfrac{4}{3}\right)\omega^{1/2}\dfrac{\sech(2\pi g(\omega,a))}{\abs{\Gamma(-\frac{1}{4}+i\,g(\omega,a))}^2}\exp[-\pi g(\omega,a)]+i\, \mathcal{G}(a,\omega).
\end{equation}
In contrast with the previous case, the conductivity in this scenario contains an explicit dependency on the rotation parameter $a$. This naturally prompts the question of how rotation influences the AC conductivity, as illustrated in Figure~\ref{fig:case2}. To address this, we begin by examining the non-rotating case, where $a=0$, represented by the black curves. In this case, and in accordance with \eqref{eq:sigma2}, the real part of the conductivity again features a delta function located at $\omega=0$. This arises from the behavior of the denominator $\abs{\Gamma(-\frac{1}{4}+i\,g(\omega,a))}^2$, which vanishes more rapidly than the numerator as $\omega \to 0$. 
 In addition, the left panel of Figure~\ref{fig:case2} reveals the presence of an apparent gap in the low-frequency regime of $\text{Re}[\sigma]$. We will come back to this matter further down; for now, we denote the width of the gap by $\omega_{g}$. 
 For frequencies above this gap, $\omega>\omega_g$, the function $\omega^{1/2}$ dominates the behavior of $\text{Re}[\sigma]$, establishing a monotonically increasing asymptotic trend growing slower than the isotropic case of Figure~\ref{fig:case1}.
 
When the rotation is switched on ($a\neq 0$), the asymptotic behavior of $\text{Re}[\sigma]$ undergoes a notable transformation. In this scenario, the Dirac delta function at $\omega=0$ remains, but it is now accompanied by the appearance of a narrow gap at low frequencies. This discernible gap can be characterized by noting that the values assumed by $\text{Re}[\sigma]$ within the interval $0<\omega \le \omega_{g}$ are significantly smaller than those beyond this range, $\omega>\omega_{g}$. As a result, within the region $0<\omega \le \omega_{g}$, we can approximate $\text{Re}[\sigma]$ effectively as zero. Under this assumption, we can estimate the width of the gap as a function of the rotation parameter, leading to
\begin{equation}\label{eq:gap_c2}
\omega_{g}/T_c \approx 8\qty(\dfrac{a}{\ell})^2+10.5,
\end{equation}
which fits very well for small values of the rotation, particularly within $0 \leq a \leq 2/5$. Moreover, this equation indicates that the gap persists even in the absence of rotation, as expected from our earlier discussion. We note that the increase of the frequency gap with the rotation parameter indicates that, once the condensate is formed, the superconducting state becomes more robust against excitations, as a larger amount of energy is required to break the Cooper pairs. In this sense, the system departs from the fundamental BCS prediction ($\omega_g/T_c \approx 3.5 $) and moves into a strong–coupling regime reminiscent of high-$T_c$ superconductivity.

It is interesting to notice that, when $\omega>\omega_{g}$, with the rotation switched on ($a\neq 0$), the curve corresponding to $\text{Re}[\sigma]$ exhibits a peak, beyond which it undergoes an exponential fall. In this high frequency regime, the term $a^2 \omega/4 \ell \Xi$ dominates over $\expval{\mathcal{O}}/\ell \omega$ in the function $g(\omega, a)$, leading to an asymptotic behavior of the form 
\begin{equation}\label{eq:Asymp_Sigc2}
\text{Re}[\sigma]\sim \omega^2 \exp\left(-2\pi a^2 \omega/4\ell \Xi \right),
\end{equation}
which explicitly reflects the dependence on the rotation parameter. 
Therefore, the combined effect of rotation and anisotropy in the background is accountable for inducing the peak and vanishing exponential behavior. This is a noteworthy characteristic in our model, as it closely resembles the behavior reported in both theoretical and experimental studies of high-$T_c$ superconductivity (see, for instance: \cite{ZIMMERMANN199199,PhysRevB.54.700,Dressel2013-mq,Michon2023-lq,Boyack2023-yu}). 

On the other hand, regarding the imaginary part $\text{Im}[\sigma]$, the functional dependence on the frequency is simpler as it is defined by the function $\mathcal{G}(a,\omega)$ in \ref{eq:func2}. At low frequencies, it follows the inverse behavior $\text{Im}[\sigma] \sim 1/\omega$, while beyond a certain minimum ---occurring at
\begin{equation}\label{eq:Imsig2_min}
\omega_{\text{min}} \approx  \dfrac{2\expval{\mathcal{O}}}{\ell a}, 
\end{equation}
for $a\neq0$, it transits into a linear regime characterized by $\text{Im}[\sigma] \sim \omega$. Nonetheless, as expected from equation \eqref{eq:func2}, such a minimum does not appear in the non-rotating case ($a=0$). In this case, the function $\text{Im}[\sigma]$ approaches to a constant in the high frequency limit, as shown by the black curve in the right panel of Figure~\ref{fig:case2}. 

Regarding the present analysis, note that our analytical results (solid lines) show excellent agreement with the numerical computations (dashed lines). However, this correspondence remains accurate primarily for small values of the rotation parameter, typically within the interval $0 \leq a \leq 2/5$. This is because the functions $\mathcal{F}(a)$ and $\mathcal{G}(a, \omega)$ were specifically chosen to match the analytic and numerical results within this range of values for $a$.

\begin{figure}[tbp]
\centering
\includegraphics[width=0.495\textwidth]{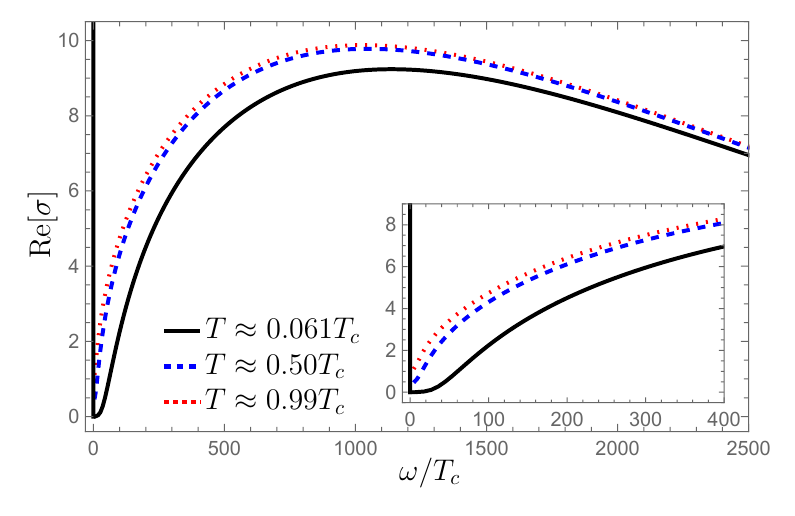}
\includegraphics[width=0.495\textwidth]{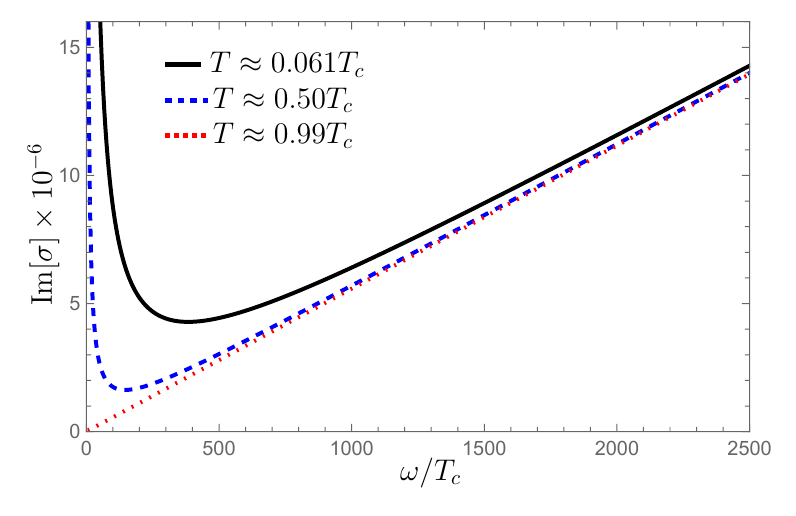}
\caption{Real and imaginary parts of the conductivity as a function of the frequency for different temperatures. For this plot, we choose $a=1/20$, $\ell =1$ and $r_h=1$.}
\label{fig:case3}
\end{figure}

To conclude this section, let us examine the temperature dependence of the conductivity. From Figure~\ref{fig:case3}, we observe that as the temperature increases, the real part of the conductivity shifts upward, while the imaginary part shifts downward. Moreover, the apparent gap in $\text{Re}[\sigma]$ observed at low temperatures gradually closes and eventually vanishes at high temperatures. A plausible suggestion is that the estimated gap \eqref{eq:gap_c2} is, in general, a function of both temperature and rotation, since it is determined by the condensate \( \expval{\mathcal{O}} \), which itself varies with these parameters.

\section{Turning on a uniform external magnetic field}
\label{sec:Vortex}

This section is devoted to studying the transformation of the vortex lattice induced by an external magnetic field in our holographic superconductor. In doing so, we offer a comprehensive derivation and revision of fundamental aspects concerning the introduction of a uniform external magnetic field and the construction of the vortex lattice model formerly presented in \cite{Herrera-Mendoza:2022whz}. To maintain a clear contrast with earlier work, we will adhere to the notation introduced in \cite{Herrera-Mendoza:2022whz} throughout subsequent sections. Additionally, as a novel application of this theoretical framework, we successfully replicate the experimental observations of a LiFeAs superconductor.

Type II superconductors are characterized by two critical magnetic fields: the lower one, $B_{c1}$, and the upper one, $B_{c2}$. Below $B_{c1}$, perfect diamagnetism is observed. Beyond this threshold, vorticity emerges when the external magnetic field exceeds $B > B_{c1}$, and the formation of a vortex lattice becomes favorable as the field approaches the upper critical value, $B_{c2}$, from below.

The emergence of the vortex lattice stems from the minimization of the free energy at magnetic field magnitudes below the upper critical value, $B_{c2}$, as initially observed in \cite{Maeda:2009vf}. This approach involves inducing the condensation of the scalar field $\Psi$ by varying the external magnetic field perpendicular to the Lifshitz boundary ($u=0$) while keeping the temperature and chemical potential constant. To address this phenomenon, we must solve the system of equations \eqref{eq:GenFieldEqns_rot} in the vicinity of the upper critical magnetic field, using a perturbation parameter, $\epsilon = (B_{c2}-B)/B\ll 1$. The series expansion of the fields for this purpose is as follows
\begin{subequations}\label{eq:epsilon_expan}
	\begin{align} 
		\Psi(u,\vec{y}) &= \epsilon^{1/2} \Psi_1(u,\vec{y})+\epsilon^{3/2} \Psi_2(u,\vec{y})+\ldots, \label{eq:expan_Psi} \\ 
		A_{\mu}(u,\vec{y}) &=  A_{\mu}^{(0)}(u,\vec{y})+\epsilon A_{\mu}^{(1)}(u,\vec{y})+\ldots, \label{eq:expan_A}
	\end{align}
\end{subequations}
Notice that the spacetime indices $\mu= (t,\phi,x,y)$ run on the boundary coordinates.

In refining the ansatz, we can still impose desired physical conditions. In our case, we seek regularity and periodicity of the condensate, along with a constant magnetic field normal to the boundary and a constant chemical potential also at the boundary. In terms of the field components we implement these conditions as
\begin{align}
\Psi_{1}(u,\vec{y})=\Phi(u,y)\text{e}^{i p x},\quad A_{x}^{(0)}(u,\vec{y})=B\,y, \quad A_{t}^{(0)}(u,\vec{y})=A_{t}^{(0)}(u), \quad A_{\phi}^{(0)}(u,\vec{y})=A_{\phi}^{(0)}(u)
\end{align}

By replacing the expansions \eqref{eq:epsilon_expan} in the field equations \eqref{eq:gen_At_eq}-\eqref{eq:gen_Ax_Ay_eq}, we obtain the non-trivial zeroth-order Maxwell equations
\begin{subequations}\label{eq:zeroth_maxwell}
	\begin{align} 
		{A_t^{(0)}}''-\Biggl[\qty(\Xi^2-1)\dfrac{f'}{f} -\dfrac{\Big(2(z-1)\Xi^2-z\Big)}{u}\Biggr]{A_{t}^{(0)}}'-\dfrac{a\Xi}{\ell^2}\Biggl[\dfrac{f'}{f}-\dfrac{2(z-1)}{u}\Biggr]{A_{\phi}^{(0)}}'&=0 \label{eq:zeroth_maxwell0},\\ 
		{A_{\phi}^{(0)}}''+\Biggl[\Xi^2\dfrac{f'}{f} -\dfrac{\Big(2(z-1)\Xi^2+2-z\Big)}{u}\Biggr]{A_{\phi}^{(0)}}'+a\Xi\Biggl[\dfrac{f'}{f}-\dfrac{2(z-1)}{u}\Biggr]{A_{t.}^{(0)}}'&=0\label{eq:zeroth_maxwell1},
	\end{align}
\end{subequations}
The rest of the equations straightforwardly yield $A_{y}^{(0)}(u,\vec{y})=0$. Notice that there is no mixing between the scalar and the Maxwell fields at this order, facilitating the analytical solution 
\begin{equation}\label{eq:zeroth_max_sols}
	A_{t}^{(0)}(u) =  \mu - \rho \qty(\dfrac{u}{r_{h}})^{3-z},\qquad  A_{\phi}^{(0)}(u) =  \nu- \zeta\qty(\dfrac{u}{r_{h}})^{3-z},
\end{equation}
with $\zeta= -\dfrac{a}{\Xi}\, \rho$. Regular boundary conditions—both asymptotically and at the horizon—constrain the parameters in the following manner: the critical exponent must lie within the interval $1\leq z<3$, while the constants are fixed as $\rho = r_{h}^{3-z}\mu$ and $\zeta = r_{h}^{3-z}\nu$.

On the other hand, the zeroth-order scalar equation associated with \eqref{eq:gen_scalar_eq} is 
	\begin{align}
	 \Biggl[ &u^{z+2}\partial_{u}\Biggl(\dfrac{f}{u^{z+2}}\partial_{u}\Biggr)+\qty(\dfrac{\ell u}{r_h})^{2(z-1)}\dfrac{(\ell^2\Xi A_t^{(0)}+aA_\phi^{(0)})^2}{r_{h}^2f}-\ell^2\Big(\dfrac{ m^2}{u^2}+\dfrac{(aA_t^{(0)}+\Xi A_\phi^{(0)})^2}{r_{h}^2}\Big) \Biggr] \Phi \nonumber\\ &=\qty(\dfrac{\ell^2}{r_h})^2\Biggl[-\partial^2_{yy} 
		+(B y-p)^2\Biggr]\Phi.
	\end{align}
One can show that the product separable ansatz $\Phi(u,y) = R_n(u) \gamma_n(y;p)$ leads to two integrable differential equations. In fact, it is widely known that one of the equations results in a one-dimensional Schr\"odinger-like equation 
	\begin{align}
		\Bigl(-\partial^2_{YY}+\dfrac{Y^2}{4}\Bigr) \gamma_{n}(y;p) =& \dfrac{\lambda_n}{2} \gamma_{n}(y;p), \label{eq:spatial}
	\end{align}
where the auxiliary variable $Y:=\sqrt{2B}\qty(y-p/B)$ was introduced. Equation \eqref{eq:spatial} dictates the spatial transversal distribution of the order parameter and its solution is determined in terms of the Hermite polynomials as follows
\begin{equation}\label{eq:droplet_sol}
	\gamma_{n}(y;p) =   \text{e}^{-Y^2/4} H_{n}(Y),
\end{equation}
together with the restriction on the eigenvalues  $\lambda_{n}= 2n+1$, with $n\in \mathbb{Z}_{\geq0}$. 

The remaining piece to be determined is often referred to as the radial part of $\Phi$. The corresponding radial equation takes the form
	\begin{align}
		R_{n}''(u)+\Bigl(\dfrac{f'}{f}-\dfrac{z+2}{u}\Bigr) R_{n}'(u)=&\Biggl[-\qty(\dfrac{\ell u}{r_h})^{2(z-1)}\dfrac{(\ell^2\Xi A_t^{(0)}+aA_\phi^{(0)})^2}{r_{h}^2f^2} \nonumber\\ &+\dfrac{\ell^2}{f}\Big(\dfrac{ m^2}{u^2}+\dfrac{(aA_t^{(0)}+\Xi A_\phi^{(0)})^2}{r_{h}^2}\Big)+\qty(\dfrac{\ell^2}{r_h})^2 \dfrac{B \lambda_n}{f} \Biggr]R_{n}(u)\label{eq:radial}.
	\end{align}
This equation describes the superconducting phase transition, and its solution requires a more detailed elaboration, which will be provided in the next section.

\subsection{The external magnetic field}
\label{sub:mag_field_c}

Upon a brief inspection of Eq.~\eqref{eq:radial}, it becomes evident that the external magnetic field plays a significant role in the dynamics of the radial mode. As we explore this effect further, the requirement of a consistent solution leads to a nontrivial relation between the upper critical magnetic field and the black hole parameters.

According to the standard holographic prescription, the radial component of the scalar field near the boundary admits the asymptotic expansion
\begin{equation}\label{eq:asym_rho}
R(u) = J_{-}u^{\Delta_{-}}+J_{+}u^{\Delta_{+}},
\end{equation}
where the scaling dimensions are given by $\Delta_{\pm} = \frac{1}{2} \left( z + 3 \pm \sqrt{(z+3)^2 + 4\ell^2 m^2} \right)$. Without loss of generality, we impose the boundary condition $J_{+} = 0$, identifying $J_{-} = J$ with scaling dimension $\Delta_{-} = \Delta$.

\begin{figure}[tbp]
		\centering
\includegraphics[width=0.495\textwidth]{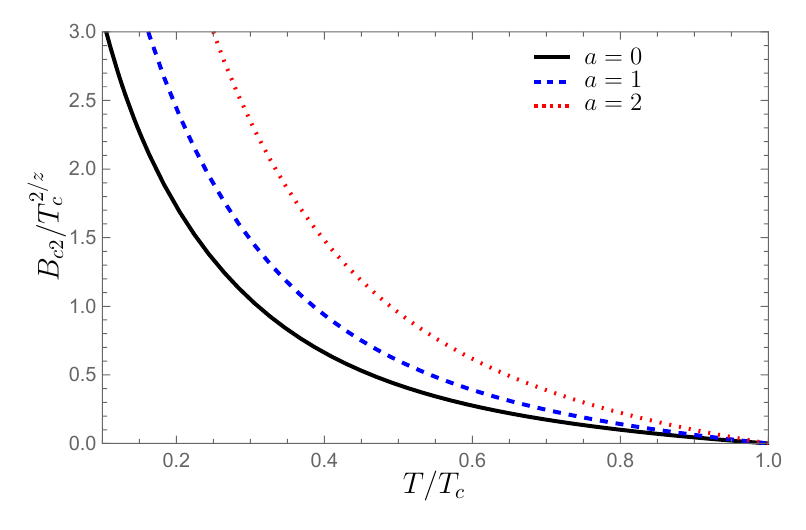}
\includegraphics[width=0.495\textwidth]{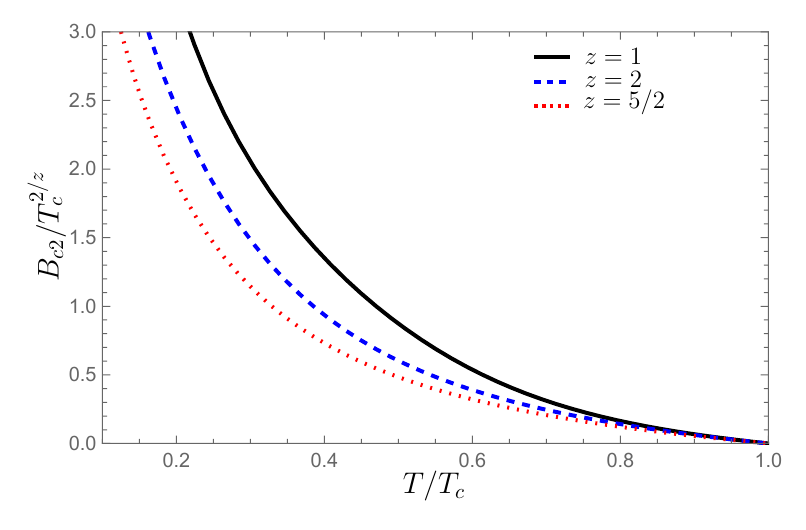}  
\caption{Upper critical magnetic field as function of temperature. The left panel shows the behavior of the upper critical magnetic field for fixed parameters $\Delta = 1$, $z = 2$, $\ell = 1$, and $u_m = 0.8$ while adopting different values of the rotation parameter $a$. In contrast, the right panel displays the corresponding behavior for $\Delta = 1$, $a = 1$, $\ell=1$, and $u_m = 0.8$ considering different values of the critical exponent $z$.
}
\label{fig:BC}
\end{figure}

On the other hand, we propose a near-horizon Taylor expansion of $R(u)$ around $u=1$
\begin{equation}\label{eq:rho_expansion}
	R(u) = R(1)+R'(1) (u-1)+\dfrac{1}{2}R''(1)(u-1)^2+\ldots\ .
\end{equation}
By evaluating the radial equation~\eqref{eq:radial} near the horizon, the expansion coefficients can be algebraically determined in terms of $R(1)$. The first two coefficients take the form
\begin{subequations}
	\begin{align}
		R'(1)=& -\dfrac{\ell^2 \qty(m^2 r_{h}^2+\ell^2 B)}{r_{h}^2 (z+3)}R(1),\\
		R''(1)=& \dfrac{1}{2(z+3)^2}\Biggl\{-\qty(\dfrac{\ell}{\Xi})^2 \qty(\dfrac{\ell}{r_h})^{2z} \left[{A_t^{(0)}}'(1)\right]^2+\ell^2 m^2\Big[\ell^2 m^2+2(z+3)\Big]\nonumber\\&+\dfrac{\ell^{4} B}{r_{h}^2}\Big[2\ell^2 m^2+ \dfrac{\ell^4 B}{r_h^2}\Big] \Biggr\}R(1).	
	\end{align}
\end{subequations} 
Once again, the matching method proves essential. Ensuring continuity and regularity requires a smooth matching between \eqref{eq:asym_rho} and \eqref{eq:rho_expansion}, and their derivatives, within a specific region of the bulk domain, say $u=u_{m}$. This matching can be achieved at the expense of the following algebraic constraints
\begin{subequations} \label{eq:match_con}
	\begin{align}	
		 J u_m^{\Delta} =& \dfrac{r_{h}^2 (z+3)+\ell^2\qty( m^2 r_{h}^2+\ell^2 B)}{r_{h}^2 (z+3)}R(1)-\dfrac{\ell^2\qty(m^2 r_{h}^2+\ell^2 B)}{r_{h}^2 (z+3)}R(1)\,u_m +\dfrac{1}{4(z+3)^2} \nonumber \\& \times \Biggl\{\ell^2 m^2\Big[\ell^2 m^2+2(z+3)\Big]+\dfrac{\ell^4 B}{r_{h}^2}\Big[2\ell^2 m^2+\dfrac{\ell^4 B}{r_h^2}\Big]-\qty(\dfrac{\ell}{\Xi})^2 \qty(\dfrac{\ell}{r_h})^{2z}[{A_t^{(0)}}'(1)]^2\Biggr\} \nonumber
		 \\ &\times\,R(1)(u_m-1)^2,
		 \\
		J \Delta u_m^{\Delta-1} =&-\dfrac{\ell^2\qty( m^2 r_{h}^2+\ell^2 B)}{r_{h}^2 (z+3)}R(1) +\dfrac{1}{2(z+3)^2}\Biggl\{-\qty(\dfrac{\ell}{\Xi})^2 \qty(\dfrac{\ell}{r_h})^{2z} [{A_t^{(0)}}'(1)]^2+\ell^2 m^2\Big[\ell^2 m^2\nonumber \\&+2(z+3)\Big]+\dfrac{\ell^4 B}{r_{h}^2}\Big[2\ell^2 m^2+\dfrac{\ell^4 B}{r_h^2}\Big] \Biggr\}R(1)(u_m-1).
	\end{align}
\end{subequations}
A particularly important implication emerges from evaluating \eqref{eq:match_con} near the upper critical magnetic field, \( B \approx B_{c2} \), in accordance with the perturbative assumption \( \epsilon \ll 1 \). By substituting the expression for \( A_t^{(0)}{}'(1) = (z - 3)\rho / r_h^{3 - z} \), obtained from ~\eqref{eq:zeroth_max_sols}, we are left with a consistency condition. This condition can be interpreted as an analytical expression for the upper critical magnetic field in terms of the critical temperature, namely,
\begin{equation}\label{eq:B_critical2}
	B_{c2}=\qty(\dfrac{\ell^{z+1}\Xi}{z+3})^{2/z}\dfrac{T^{2/z}}{\ell^4 \eta}\Biggl\{ -\delta+\Biggl[\chi+\qty(\delta^2-\chi)\qty(\dfrac{T_{c}}{T})^{6/z}\Biggr]^{1/2}\Biggr\},
\end{equation}
where the constants
\begin{subequations}
	\begin{align}
		\eta=&(1-u_m)\Bigl[(2-\Delta)u_m+\Delta\Bigr],\\
		\chi = 2&(z+3)\Big( 2(z+3)u_m^2-\ell^2m^2\eta^2\Big),\\
		\delta= 2u_m\Bigl[(z+3)+\ell^2m^2(1&-u_m) \Bigr]+\Delta (1-u_m)\Bigl[2(z+3)+\ell^2m^2(1-u_m) \Bigr] ,
	\end{align}
\end{subequations}
introduce alleviated notation. The critical temperature is given by
\begin{equation}\label{eq:T_critical}
	T_{c} = \dfrac{z+3}{\ell^{z+1} \Xi}\dfrac{1}{\qty(\delta^2-\chi)^{z/6}}\qty[\eta^2 \ell^{2z} \qty(z-3)^2 \qty(\dfrac{\ell}{\Xi})^2 \rho^2]^{z/6}.
\end{equation}
A quick check-up reveals that $B_{c2}$ goes as predicted by the Ginzburg-Landau theory near the critical temperature, with $B_{c2} \propto (1-T/T_c)$. This behavior is illustrated in Figure~\ref{fig:BC}, which also shows the effects of rotation and anisotropy on \( B_{c2} \), with the scaling dimension set to $\Delta=1$. As shown in the figure, increasing the dynamical exponent \( z \) leads to a decrease in the critical magnetic field, suggesting that anisotropic scaling makes the system more susceptible to the suppression of superconductivity. In contrast, the rotation parameter contributes to an increase in the critical magnetic field, implying that a stronger external magnetic field is required to destroy the superconducting phase.

\subsection{The vortex configuration}
\label{sub:vortex_conf}
Let us  now transition to a more phenomenological setting, aimed at modeling vortex lattice deformations in experimentally realizable superconductors. In particular, we explore whether the anisotropic holographic setup developed above can reproduce lattice transitions observed in materials such as LiFeAs, following the investigations carried out in previous works \cite{Maeda:2009vf,Guo:2014wca,Herrera-Mendoza:2022whz}.

Our construction is founded upon the most stable modes of the scalar field, which are the lowest (zero) orders, as illustrated in \cite{Maeda:2009vf}. The transversal part contributes then with
\begin{equation}
	\gamma_0(y;p_k) = \exp[-\dfrac{1}{2r_{0}^2}(y-p_kr_0^2)^2],
\end{equation}
where $r_0=1/\sqrt{B}$. The full lattice, however, requires a linear superposition over all the $p_k$ parameters 
\begin{equation}\label{eq:lc_droplet}
	\Psi_1(u,\vec{y}) = R_{0}(u) \gamma_{L}(\vec{y}),\qquad \text{with}\qquad \gamma_{L}(\vec{y}) := \sum_{k=-\infty}^{\infty} c_{k} e^{ip_{k}x}\gamma_0(y;p_k).
\end{equation}
Similarly, $R_{0}(u)$ stands for the lowest order solution to the radial equation \eqref{eq:radial}.

The lattice is characterized by the periodic occurrence of fundamental cells. However, notice from \eqref{eq:lc_droplet} that this function is already periodic along the $x$ direction, but not necessarily along the $y$ direction. Full periodicity can be introduced by 
means of the reparametrization
\begin{equation}\label{eq:parameters}
	c_{k}\equiv e^{iq_k},\qquad q_{k}\equiv \alpha k^2,\qquad p_{k}\equiv \beta k,
\end{equation}
in terms of the arbitrary constants $\alpha$ and $\beta$. Using this definition, we can express $\gamma_L$ in terms of the elliptic theta function $\vartheta_3$ as follows
\begin{equation}
	\gamma_{L}(\vec{y}) = e^{-y^2/2r_{0}^2} \vartheta_{3}(v,\tau).
\end{equation}
Above, new convenient coordinate and parameter were introduced
\begin{equation}
	v:= \dfrac{\beta}{2\pi}\qty(x-iy),\qquad \tau := \dfrac{1}{2\pi}\qty(2\alpha+i\beta^2r_{0}^2). 
\end{equation}
The $\gamma_{L}$ function inherits the symmetry properties of $\vartheta_3$. It is thus not difficult to show that $\gamma_{L}$ exhibits pseudo-periodicity in the directions characterized by the vectors
\begin{subequations}\label{eq:fund_vectors}
	\begin{align}
		\vec{b}_1 =& \dfrac{2\pi}{\beta} \partial_x,\\
		\vec{b}_2 =& \dfrac{2\alpha}{\beta} \partial_x -\beta r_{0}^2 \partial_y.
	\end{align}
\end{subequations}
Accordingly, the explicit transformations manifesting the aforementioned pseudo-periodicity are
\begin{subequations}\label{eq:psi1_prop}
	\begin{align}
		\gamma_{L}(x+\dfrac{2\pi}{\beta},y) &= \gamma_{L}(x,y) ,\\  
		\gamma_{L}\Bigl(x+\dfrac{2\alpha}{\beta},y-\beta r_{0}^2\Bigr) &=\exp[-i\qty(\alpha +\beta x)]\gamma_{L}(x,y).
	\end{align}
\end{subequations} 
\begin{figure*}[b]
	\centering
	\begin{tikzpicture}[scale=5]
		\node [black] at (0.352,0.609) {\textbullet};
		\node [black] at (0.71,0.254) {\textbullet};
		\node [black] at (0.0,0.0) {\textbullet};
		\draw (0.12, 0.35) node {$\vec{b}_{1}$}; 
		\draw (0.35, 0.05) node {$\vec{b}_{2}$}; 
		\draw[black,-latex] (0cm,0cm) coordinate(O) -- (20:0.75cm) coordinate (r) node[pos=1.02,anchor=west]{};
		\draw[black,-latex] (0cm,0cm) -- (60:0.7cm) coordinate (i) node[pos=1.02,anchor=west]{};
		\draw pic ["$\theta$",angle eccentricity=1.33,draw,-latex,angle radius=1cm] 
		{angle = r--O--i};
	\end{tikzpicture}    
	\caption{Fundamental vectors of the vortex lattice associated with the type-II holographic superconductor.}
	\label{fig:fund_vectors}
\end{figure*}

The lattice structure is fully attained up to this point. It is straightforward to verify that the modulus squared $\abs{\gamma_L}^2$ is already periodic while fundamental regions conforming the lattice are generated by the vectors  \eqref{eq:fund_vectors}. Consequently, any point on the lattice can be expressed as a linear combination of $\vec{b}_1$ and $\vec{b}_2$. In agreement with  Ginzburg-Landau theory, here the magnetic flux is quantized as $B\times 2\pi r_{0}^2 = 2\pi$.

The zeros of $\abs{\gamma_{L}} $ indicate the locations of vortex cores, where the order parameter vanishes ($\expval{\mathcal{O}}\sim \abs{\gamma_{L}} = 0$). These zeros can be expressed in a compact form as given in the following equation \cite{Maeda:2009vf}
\begin{equation}\label{eq:gamma_zeros}
	\vec{x}_{n_1,n_2} =\qty(n_1+\dfrac{1}{2})\vec{b}_1+ \qty(n_2+\dfrac{1}{2})\vec{b}_2,
\end{equation}
for integers $n_1$ and $n_2$. 
Now, using equations \eqref{eq:fund_vectors} and \eqref{eq:gamma_zeros}, we can establish the following relationships between two adjacent vectors of equal magnitude in the fundamental cell
\begin{equation}\label{eq:vortex_pos}
		\cos{\theta} =\sqrt{1-\dfrac{\beta^4}{4\pi^2 B^2}}, \qquad \alpha = \pi \cos{\theta},
\end{equation}
wherein the angle between adjacent vectors is denoted by $\theta$, as shown in Figure \ref{fig:fund_vectors}. In this manner, relations \eqref{eq:vortex_pos} provide a mechanism for continuously deforming the lattice structure by virtue of the external magnetic field. 
It is noteworthy that the parameter $\beta$ remains arbitrary and can be appropriately chosen as a function of the external magnetic field. Specifically, we can select $\beta$ in such a way that the vortex lattice deforms in a desired manner as the external magnetic field is applied. The specific function that relates $\beta$ to the magnetic field is expected to depend on the properties of the material as elaborated in the following discussion on a experimental realization.


\begin{figure}[t!]
\centering
\begin{subfigure}[b]{.32\textwidth}
	\centering
	\includegraphics[width=\textwidth]{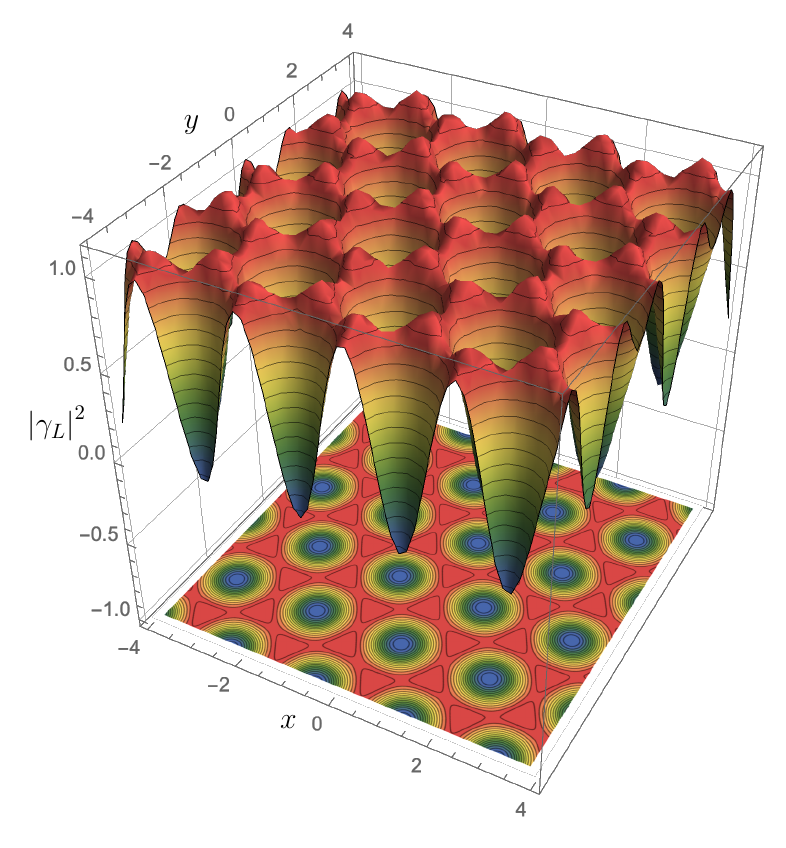}
	\caption{}
	\label{fig:tr_lat2}
\end{subfigure}
\begin{subfigure}[b]{.32\textwidth}
\centering
\includegraphics[width=\textwidth]{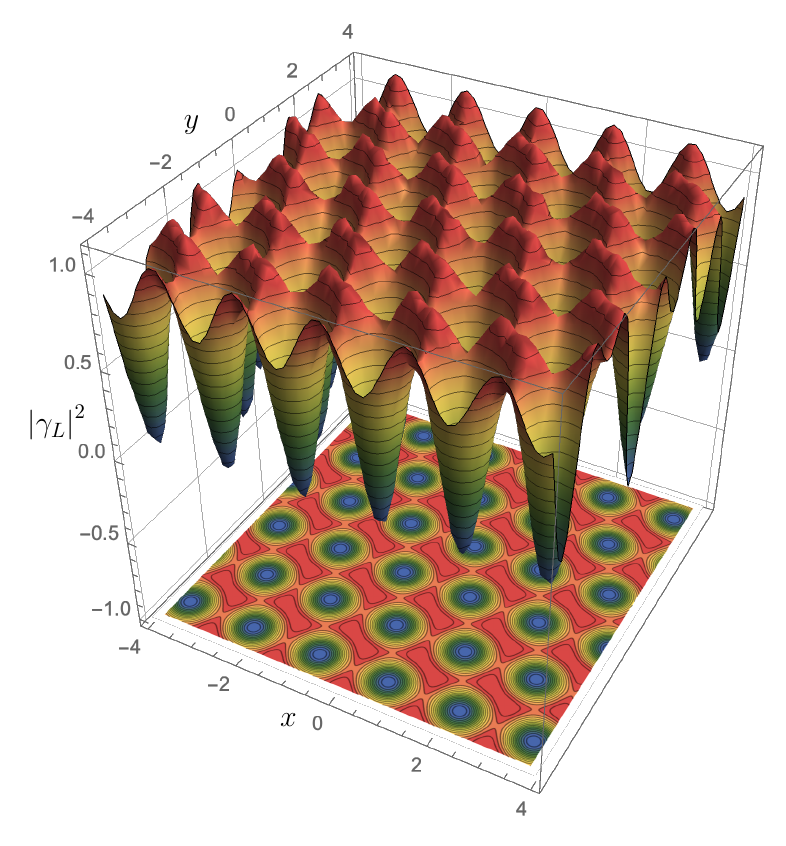}  
\caption{}  
\label{fig:rb1_lat}
\end{subfigure}  
\begin{subfigure}[b]{.32\textwidth}
\centering
\includegraphics[width=\textwidth]{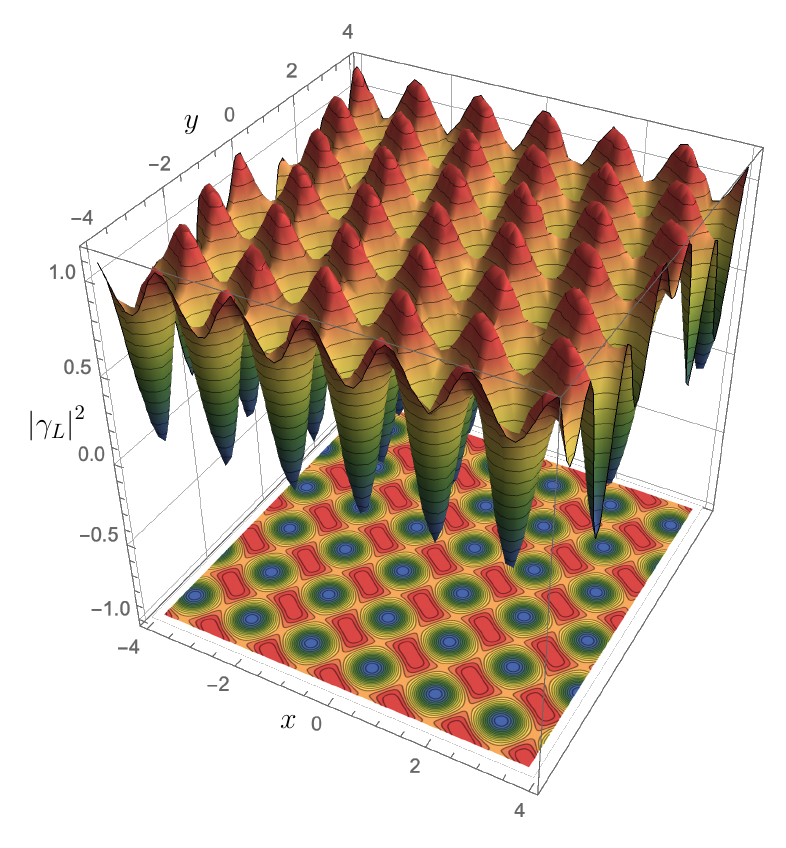}  
\caption{}  
\label{fig:rb2_lat}
\end{subfigure}
\begin{subfigure}[b]{.32\textwidth}
\centering
\includegraphics[width=\textwidth]{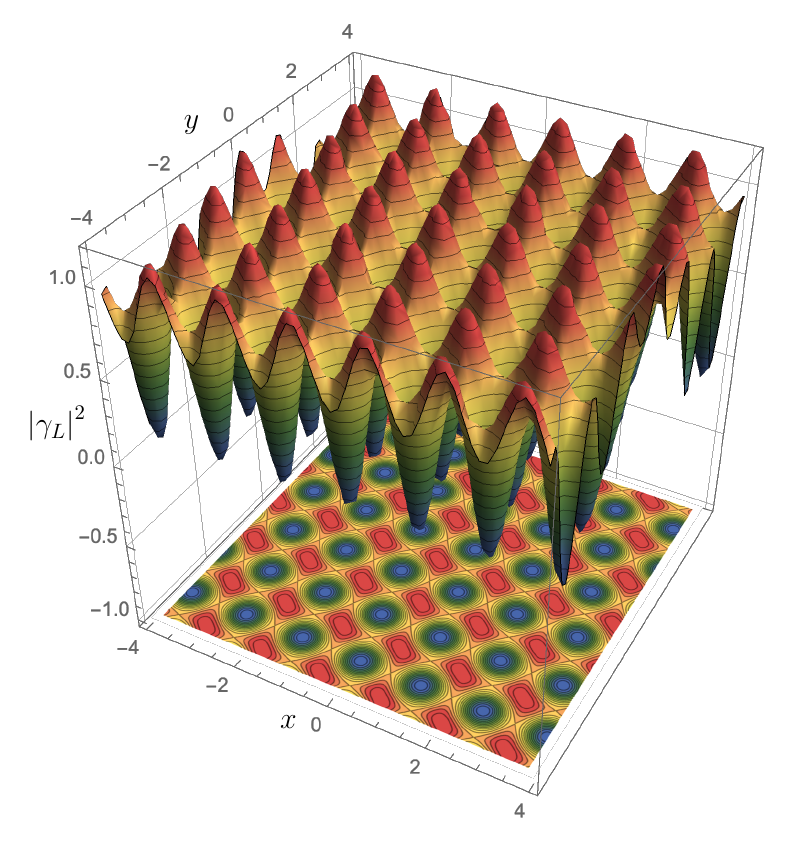}
\caption{}
\label{fig:rb3_lat}
\end{subfigure}
\begin{subfigure}[b]{.32\textwidth}
\centering
\includegraphics[width=\textwidth]{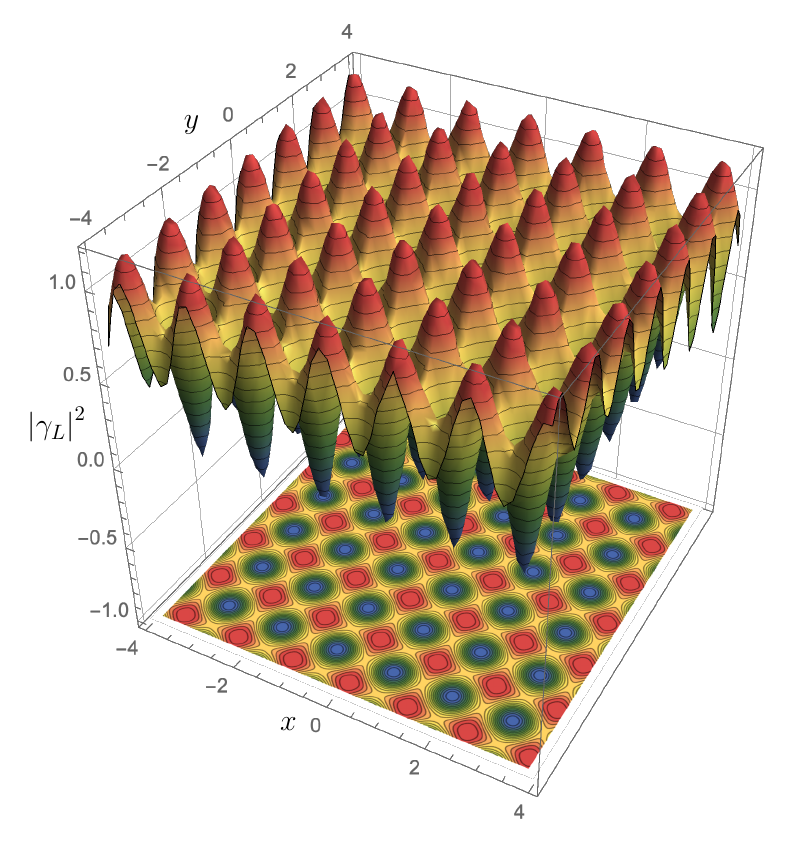}
\caption{}
\label{fig:sq2_lat}
\end{subfigure}
\caption{The transformation of the LiFeAs superconductor vortex lattice from the triangular (a) at $2$ T, passing the intermediate states (b)-(d) at 3, 3.5, and 3.75 T respectively, to the square lattice (e) at 4 T,  when increasing external magnetic field. }
\label{fig:lat_def2}
\end{figure}

\subsubsection*{Vortex lattice deformation of a lithium-iron-arsenide (LiFeAs) superconductor }

As we already remarked, the relations \eqref{eq:vortex_pos} define continuous deformations of a given vortex lattice driven by the external magnetic field. Here we want to study how these deformations take place when starting from a particular vortex configuration.
In particular, we would like to go beyond the theoretical scopes of the model presented here and qualitatively describe some experimental behavior concerning the vortex lattice transformation of iron-based superconductors. We recall that the description of a FeSe compound was recently detailed in \citep{Herrera-Mendoza:2022whz}.

An interesting iron-based superconductor showing experimental vortex lattice transformations is the LiFeAs superconductor explored in \cite{Zhang:2019prb}. Unlike the FeSe case, here the transformation occurs in a slightly different regime of the external magnetic field but evolving in the same direction, from a triangular to a square lattice. Concretely, the transformation occurs from a triangular lattice at $2$ T, passing through intermediate configurations at $3$ T, $3.5$ T, and $3.75$ T and ending with a square structure at $4$ T. In order to model such a behavior, the $\beta$ parameter must be considered linear in $B$, i.e.
\begin{equation}\label{eq:beta_1}
\beta= a_{1}+a_{2}\,B, \qquad 2\ \text{T}  < B< 4\ \text{T},
\end{equation}
with coefficients $a_1=1.57161\ \text{T}^{1/2}$ and $a_2=0.85991\ \text{T}^{-1/2}$, while it is important to bear in mind that the temperature is kept fixed in the experiment. The behavior reproduced by our holographic model is illustrated in Figure \ref{fig:lat_def2}.

Here it is worth noticing that the linear behavior that the $\beta$ parameter presents with respect to the external magnetic field $B$ in the LiFeAs superconductor, contrasts with the quadratic dependence that this parameter displays for the FeSe superconducting material \cite{Herrera-Mendoza:2022whz}. This difference indicates that the dependence of this parameter on the external magnetic field $\beta(B)$ is determined by the intrinsic properties of diverse superconductors.




\section{Conclusions and discussion}
We have developed a holographic model for a type-II s-wave superconductor that incorporates rotating and anisotropic aspects in the black hole background. Using both analytical and numerical approaches, we examined how rotation and anisotropy influence the properties of the superconductor model. Concerning the condensation phenomena, our findings reveal that an increment (or decrement) in the rotation leads to a corresponding reduction (or increment) in the magnitude of the condensation curve, as illustrated in Figure \ref{fig:Condensate}. This scaling effect was qualitatively reproduced by the analytical approach, in which the matching method was used, and supported by the numerical approach using the shooting method.
Regarding the results for the AC conductivity, we worked out two cases, the isotropic ($z=1$, $\Delta=1$) and the anisotropic case ($z=2$, $\Delta=1$), where novel closed formulas for the conductivity were obtained. The isotropic case remained unaffected by the presence of a nontrivial rotation, indicating that to study the effect induced by this quantity, we must take into account the anisotropic properties of the background. In this regard, the interplay between rotation and anisotropy brought forth a noteworthy behavior not reported in previous holographic models. Concretely, our results have shown that the rotation induces a peak in the low-frequency region of $\text{Re}[\sigma]$, followed by an exponentially vanishing tendency at higher frequencies. Moreover, our findings also revealed that the magnitude of the peak falls monotonically and is shifted to a lower frequency as the rotation increases, as shown in the left panel of Figure~\ref{fig:case2}.    
Altogether, this behavior aligns with predictions from high-temperature superconductor models and experiments \cite{ZIMMERMANN199199,PhysRevB.54.700,Dressel2013-mq,Michon2023-lq,Boyack2023-yu}. Within this context, the appearance of the peak and the vanishing behavior at high frequencies are attributed to the effects of quasiparticle damping driven by impurities or defects in the superconducting system. Consequently, our results suggest an interesting connection between the rotation of the anisotropic black hole in our holographic superconductor model and quasiparticle damping in superconducting materials.
In relation to the imaginary component $\text{Im}[\sigma]$, the effect of the rotation is manifested in increasing the amplitude of the curve at higher frequencies as the rotation parameter increases, depicted in the right panel of Figure~\ref{fig:case2}.

Furthermore, we considered the superconducting system subject to a uniform magnetic field. We employed the matching method to derive an analytical expression for the upper critical magnetic field, the threshold below which the vortex phenomenon occurs. The results reveal that the upper critical magnetic field is affected by both the dynamical scaling exponent \( z \) and the rotation parameter \( a \). Specifically, increasing \( z \) diminishes the value of the critical magnetic field required to destroy the superconducting state, while increasing \( a \) enhances it, see Figure~\ref{fig:BC}. These findings imply that anisotropic scaling tends to weaken superconductivity by making it more sensitive to external magnetic fields, whereas rotation strengthens it in the sense that stronger external magnetic fields are needed to destroy the superconducting state.

To construct the Abrikosov vortex lattice, we adopted an approach inspired on previous works \cite{Maeda:2009vf,Guo:2014wca,Herrera-Mendoza:2022whz}. Our method yielded promising results as we were able to describe continuous transformations of the vortex lattice driven by the external magnetic field in a unified manner. Notice that such transitions can also be described in field theory superconductivity by modifications of the Ginzburg-Landau theory, although diverse lattice transformations arise depending on the materials and the experimental setups \cite{PhysRevLett.78.4273,PhysRevB.55.R8693,PhysRevB.99.144514}. 
Specifically, by appropriately choosing the $\beta$ parameter as a function of the external magnetic field, we successfully reproduced the vortex transformation observed in recent experiments concerning iron-based superconductors, concretely LiFeAs \cite{Zhang:2019prb}. In this case, it was sufficient to adjust the $\beta$ parameter as a linear function of the external magnetic field $B$, suggesting that the $\beta$ parameter will be defined according to the properties of the superconducting material under consideration. The freedom in choosing $\beta$ offers additional modeling possibilities that are worth investigating in detail for their potential application to other relevant findings.

As a final comment, we would like to draw attention to a noteworthy point of comparison with our model that was presented in  \cite{Xia:2021jzh}. Therein, the authors introduce a numerical approach for the construction of the vortex lattice capable of reproducing triangular and square structures driven by both temperature and the external magnetic field. It will be interesting to include the effect of both temperature and magnetic field within our analytic approach. The first idea that comes to mind consists in assuming a more general dependence of the angle between the fundamental vectors on the external magnetic field and the temperature, $\theta=\theta(B,T)$. This relation can be realized by making the parameter $\beta$ dependent on the temperature as well (think of the coefficients $a_1$ and $a_2$ depending on T, for instance). Thus, we arrive at the need to perform a series of experiments in which the vortex lattice deformation is driven by changes in both the temperature and the magnitude of the external magnetic field. The corresponding experimental data would allow us to fit the observed $\theta=\theta(B,T)$ dependence.  

\acknowledgments

All the authors are grateful to Manuel de la Cruz for enriching discussions. The authors acknowledge financial support from  FORDECYT-PRONACESCONACYT Grant No. CF-MG-2558591, CONAHCYT Grant No. A1-S-38041, and VIEP-BUAP Grant.  JAHM acknowledges support from CONAHCYT through PhD Grant No. 750974 and Postdoctoral support under project No. CF-MG-2558591. DFHB and JAMZ are also grateful to CONAHCYT for a \emph{Estancias posdoctorales por M\'{e}xico} Grant No. 372516 and project 898686, respectively. 


%
\appendix
\section{Analytical study of the condensation operator}\label{ape:A}
Here we want to find an analytic expression for the expectation value of the condensation operator in terms of the black hole parameters. To achieve this, we solve the system \eqref{eq:CondeFieldEqns} using the matching method \cite{Zhao:2013pva}, which involves determining the solutions at the horizon ($u=1$) and the asymptotic boundary ($u=0$) and matching them at an intermediate point within the bulk $0< u_{m}<1$.

To find the horizon solutions, we propose the following Taylor series expansions of the fields around $u=1$
\begin{subequations}\label{eq:HorizonExpans}
	\begin{align}
		\psi(u) &= \psi(1)+\psi'(1)\qty(u-1)+\dfrac{\psi''(1)}{2}\qty(u-1)^2+\cdots\,, \\
		A_t(u) &= A_t(1)+A_t'(1)\qty(u-1)+\dfrac{A_t''(1)}{2}\qty(u-1)^2+\cdots\,,\\
		A_\phi(u) &= A_\phi(1)+A_\phi'(1)\qty(u-1)+\dfrac{A_\phi''(1)}{2}\qty(u-1)^2+\cdots\,.
	\end{align}
\end{subequations}
In these expressions the regularity of the gauge field at the horizon leads to the boundary conditions
\begin{equation}\label{eq:HorizonBC1}
	A_t(1)=0, \qquad A_\phi(1)=0, 
\end{equation}
while $\psi(1)$ is keep finite.
The remaining coefficients are determined by making the expansions \eqref{eq:HorizonExpans} consistent with the field equations \eqref{eq:CondeFieldEqns}. After some algebraic computations, we find 
\begin{subequations}
	\begin{align}
		A_\phi'(1)=-\dfrac{a}{\Xi}A_t'(1), \qquad &\psi'(1)=-\dfrac{\ell^2m^2}{z+3}\psi(1); \\
		A_t''(1)  =-\dfrac{2\psi(1)^2+(z+3)(z-2)}{z+3}&A_t'(1), \quad A_\phi''(1)=-\dfrac{a}{\Xi}A_t''(1);\\
		\psi''(1)  =\dfrac{\ell^2m^2}{z+3}\qty[1+\dfrac{\ell^2m^2}{2(z+3)}]&\psi(1)-\dfrac{\ell^2A_t'(1)^2}{2\Xi^2(z+3)^2}\qty(\dfrac{\ell}{r_h})^{2z}\psi(1).
	\end{align}
\end{subequations}
Note that the higher-order terms are also expressed in terms of the lowest-order ones. On the other hand, the regular solutions at the asymptotic boundary read 
\begin{equation}\label{eq:BoundarySolA}
	\psi=J\, u^{\Delta}, \quad A_t=\mu-\rho\qty(\dfrac{u}{r_h})^{3-z}, \quad A_\phi=\nu-\zeta\qty(\dfrac{u}{r_h})^{3-z},
\end{equation}
where we identify $J$ as related to the expectation value of the condensation operator, $J = \expval{\mathcal{O}}/{\sqrt{2}r_h^{\Delta}}$, and $\Delta$ as the scaling dimension of the operator $\mathcal{O}$.

Thus far we have found expressions for both regimes, near horizon $(u=1)$ and asymptotic $(u=0)$ solutions.
In what follows, we require that there is an intermediate point, $u=u_m$, within the bulk, where the asymptotic solutions, \eqref{eq:BoundarySolA} and \eqref{eq:HorizonExpans}, match and their first derivatives as well. This assumption leads us to four independent relations
\begin{subequations}\label{eq:Matching}
	\begin{align}
		Ju_m^\Delta&=\psi(1)-\dfrac{\ell^2m^2}{z+3}\psi(1)(u_m-1) + \Bigg\{\dfrac{\ell^2m^2}{z+3}\qty[1+\dfrac{\ell^2m^2}{2(z+3)}]\nonumber\\ &\quad -\dfrac{\ell^2A_t'(1)^2}{2\Xi^2(z+3)^2}\qty(\dfrac{\ell}{r_h})^{2z}\Bigg\}\dfrac{\psi(1)}{2}(u_m-1)^2,\label{eq:psiMatching}\\
		J\Delta u_m^{\Delta-1}&=-\dfrac{\ell^2m^2}{z+3}\psi(1)+\Bigg\{\dfrac{\ell^2m^2}{z+3}\qty[1+\dfrac{\ell^2m^2}{2(z+3)}]\nonumber \\&\quad -\dfrac{\ell^2A_t'(1)^2}{2\Xi^2(z+3)^2}\qty(\dfrac{\ell}{r_h})^{2z}\Bigg\}\psi(1)(u_m-1),\label{eq:DpsiMatching}\\
\mu-\rho\qty(\dfrac{u_m}{r_h})^{3-z}&=A_t'(1)(u_m-1)-\dfrac{2\psi(1)^2+(z+3)(z-2)}{2(z+3)}A_t'(1)(u_m-1)^2,\label{eq:AtMatching}\\
		 -(3-z)\dfrac{\rho}{r_h}\qty(\dfrac{u_m}{r_h})^{2-z}&=A_t'(1)-\dfrac{2\psi(1)^2+(z+3)(z-2)}{z+3}A_t'(1)(u_m-1).\label{eq:DAtMatching}  
	\end{align}
\end{subequations}
From \eqref{eq:psiMatching} and \eqref{eq:DpsiMatching} we arrive at
\begin{subequations}
	\begin{align}
		J=&\dfrac{u_m^{1-\Delta}[\ell^2m^2(1-u_m)+2(z+3)]}{(z+3)[\Delta(1-u_m)+2u_m]}\psi(1) \label{eq:Jcoeff},\\
		&\qquad A_t'(1)=-\gamma\Xi\qty(\dfrac{r_h}{\ell})^z,
	\end{align}
\end{subequations}
where
\begin{equation}
	\gamma=\sqrt{\ell^2m^4+\dfrac{2\ell^2m^2(z+3)\left\{u_m\qty[u_m(\Delta-2)-4(\Delta-1)]+3\Delta\right\}+4\Delta(z+3)^2}{\ell^2\qty[\Delta(1-u_m)+2u_m](1-u_m)}}.
\end{equation}
Similarly, from \eqref{eq:DAtMatching} we obtain
\begin{equation}\label{eq:psi(1)_0}
	\psi(1)^2=\dfrac{(z+3)}{2(1-u_m)}\qty[\dfrac{\ell^z(3-z)}{\gamma\Xi r_h^3}\rho u_m^{2-z}-(z-2)(1-u_m)-1].
\end{equation}
At this point, let us recall that the black hole temperature is given by Eq.~\eqref{eq:BBTemperature} with \(d = 5\), taking the form
\begin{equation}
	T = \dfrac{1}{4\pi} \dfrac{(z+3)r_h^z}{\ell^{z+1} \Xi},
\end{equation}
which allows us to rewrite Eq.~\eqref{eq:psi(1)_0} as an explicit function of the black hole temperature
\begin{equation}\label{eq.psiHorizonT}
	\psi(1)^2 = \dfrac{(z+3)}{2(1 - u_m)} \qty[\dfrac{\ell^z (3 - z)}{\gamma\, \Xi^{1 + 3/z}} \qty(\dfrac{z+3}{4\pi \ell^{z+1}})^{3/z} \dfrac{\rho\, u_m^{2 - z}}{T^{3/z}} - (z - 2)(1 - u_m) - 1],
\end{equation}
which, together with Eq.~\eqref{eq:Jcoeff}, provides a concrete expression for the boundary coefficient \(J\) in terms of the black hole temperature.

\bibliography{Bibliography}
\bibliographystyle{./JHEP}








\end{document}